\documentclass[fleqn,usenatbib]{mnras}

\usepackage{newtxtext,newtxmath}
\usepackage[T1]{fontenc}
\usepackage{subcaption}

\DeclareRobustCommand{\VAN}[3]{#2}
\let\VANthebibliography\thebibliography
\def\thebibliography{\DeclareRobustCommand{\VAN}[3]{##3}\VANthebibliography}

\usepackage{graphicx}	
\usepackage{amsmath}	
\usepackage{xcolor}
\usepackage{soul}
\usepackage{multirow}
\usepackage[dvipsnames]{xcolor}
\usepackage{makecell}

\usepackage[table]{xcolor}
\newcommand{\pos}[1]{\cellcolor{teal!20}{#1}} 
\newcommand{\nega}[1]{\cellcolor{violet!20}{#1}} 
\definecolor{neworange}{HTML}{D55E00}
\newcommand{\orange}[1]{{\color{neworange}#1}}

\definecolor{voidcolour}{RGB}{222,235,247} 
\definecolor{wallcolour}{RGB}{158,202,225} 

\title[Disentangling modified gravity and galaxy bias with field-level inference]{Disentangling modified gravity and galaxy bias with field-level inference}

\author[]{
Sophie Hoyland\thanks{E-mail: \url{sophie.hoyland@port.ac.uk}},$^{1}$
Daniela Saadeh\thanks{E-mail: \url{daniela.saadeh@durham.ac.uk}},$^{1,2}$
Kazuya Koyama,$^{1,3,4}$
Harry Desmond$^{1}$
\\
$^{1}$Institute of Cosmology and Gravitation, University of Portsmouth, Dennis Sciama Building, Burnaby Road, Portsmouth, PO1 3FX, United Kingdom \\
$^{2}$Institute for Computational Cosmology, Department of Physics, Durham University, South Road, Durham DH1 3LE, United Kingdom \\
$^{3}$ Kavli IPMU (WPI), UTIAS, The University of Tokyo, Kashiwa, Chiba 277-8583, Japan\\
$^{4}$ Yukawa Institute for Theoretical Physics, Kyoto University, Kyoto 606-8502, Japan
}

\date{Accepted XXX. Received YYY; in original form ZZZ}
\pubyear{\the\year{}}

\begin{document}
\label{firstpage}
\pagerange{\pageref{firstpage}--\pageref{lastpage}}
\maketitle 

\begin{abstract}

We present a field-level inference framework for testing gravity with large-scale structure, exploiting the full information of the galaxy distribution. Traditional analyses based on the power spectrum discard non-Gaussian and Fourier phase information, resulting in strong degeneracies between modified gravity (MG) and galaxy bias. Our approach overcomes this limitation by performing a Bayesian analysis directly on the three-dimensional galaxy number counts, jointly constraining MG and bias parameters using both amplitudes and phases. As an illustrative application, we analyse mock data in real space in the context of $f(R)$ gravity and a non-linear galaxy bias model designed to mimic the real-data 2M++ BORG analysis of \cite{Jasche_physical_2019}.
Non-linear structure formation is modelled using COmoving Lagrangian Acceleration (COLA) under different gravity strengths, parameterised by $f_{R0}$. The resulting dark-matter fields are then mapped to mock galaxy catalogues via a non-linear bias prescription. We demonstrate that, with the initial phases assumed known, including non-Gaussian and phase information yields tighter constraints on both $f_{R0}$ and the primary bias parameter $\beta$ (corresponding to the linear galaxy bias on large scales), relative to power-spectrum-only analyses. Notably, the field-level approach breaks the degeneracies between MG and galaxy bias inherent to two-point statistics. Through a cosmic-web classification into voids, walls, filaments and clusters, we find that under-dense regions are the primary drivers in distinguishing gravity at the field level. Finally, we establish the robustness of our pipeline against variations in initial conditions, Poisson noise, and galaxy-field thresholding, providing a powerful path forward for field-level tests of gravity.
\end{abstract}

\begin{keywords}
cosmology: large-scale structure of Universe -- galaxies: statistics -- methods: data analysis
\end{keywords}

\section{Introduction}\label{sec:intro}

Field-level inference (FLI) has emerged as a powerful approach for extracting cosmological information from the non-linear regime of the large-scale structure (LSS) \citep{Jasche_bayesian_2013, Wang_ELUCID_2014, Jasche_past_2015, Lavaux_unmasking_2016, Jasche_physical_2019, Leclercq_accuracy_2021, Andrews_bayesian_2023, Ding_shear_2025}. Traditional approaches to cosmological inference typically compress the three-dimensional matter distributions into summary statistics, most commonly the power spectrum, $P(k)$, the Fourier-space counterpart of the two-point correlation function. The power spectrum provides a complete statistical description of the initial density field if it is Gaussian, as expected from simple inflationary models, because in that case the Fourier modes are independent and their phases are random. Non-linear gravitational evolution breaks this independence by coupling modes together. This generates higher-order correlations among Fourier modes, which manifest as correlations between their phases and as departures from Gaussian statistics on small and intermediate scales ($e.g.$, \citealt{Soda_nonlinear_1992, Chiang_phase_2000, Watts_universal_2003}). These correlations encode morphological information about the spatial organisation of structure—how overdensities align, merge and collapse to form the cosmic web \citep{Coles_large-scale_2001}. Such information can discriminate between different theories of gravity and models of galaxy formation, but it is not captured by two-point statistics like the power spectrum \citep{Rimes_information_2005, Carron_inadequacy_2012, Carron_information_2015}. 
In contrast, by operating directly on the three-dimensional matter or galaxy field, FLI naturally preserves the full Fourier information, both phases and amplitudes, thereby exploiting the complete non-Gaussian morphology of the cosmic web \citep{Jasche_bayesian_2013, Seljak_towards_2017} although limitations remain, including a cutoff scale set by the voxel size \footnote{We note that $n$-piont statistics are similarly subject to scale cutoffs.} and a potentially enhanced sensitivity to systematics 

Observations of the cosmic web from surveys such as the 2-degree Field Galaxy Redshift Survey (2dFGRS) \citep{Colless_2dFGRS_2001}, the Sloan Digital Sky Survey (SDSS) \citep{Tegmark_SDSS_2004} and the Two Micron All Sky Survey (2MASS) \citep{Huchra_2MASS_2012} have provided detailed three-dimensional maps of the local Universe. These datasets are now being significantly extended by Stage~IV surveys such as the Dark Energy Spectroscopic Instrument (DESI) \citep{DESI_validation_2024}, Euclid \citep{Euclid_overview_2025} and the Vera Rubin Observatory's Legacy Survey of Space and Time (LSST) \citep{LSST_LSST_2009}, which are measuring galaxy clustering and weak lensing with unprecedented depth, volume and statistical precision. The majority of the total cosmological constraining power in these surveys is estimated to reside in the non-Gaussian regime \citep{Jasche_physical_2019, Euclid_preparation_2020}, where the density field has departed from linear evolution. Fully exploiting this information requires sophisticated analysis techniques capable of operating directly on the three-dimensional field, without discarding the phase correlations essential for reconstructing the cosmic web morphology \citep{Jasche_bayesian_2013, Modi_cosmological_2018}.

FLI provides precisely this capability: rather than compressing the data, it employs physical forward modelling in which three-dimensional predictions for the full matter or galaxy field are simulated and compared voxel-by-voxel directly to the observed field \citep{Jasche_bayesian_2013, Leclercq_accuracy_2021, Boruah_map-based_2024}. Although carefully selected sets of higher-order summary statistics, such as the galaxy bispectrum \citep{Hahn_constraining_2021, Gualdi_joint_2021} and Minkowski functionals \citep{Schmalzing_Minkowski_1996, Fang_new_2017}, can recover a subset of the information encoded in non-Gaussian structures ($e.g.$, \citealt{Kilbinger_cosmological_2005}), they often struggle to capture the full hierarchical nature of the LSS \citep{Bernardeau_large-scale_2002} and have poorly-known sampling distributions, which makes inference difficult. As a result, cosmological inference based on Minkowski functionals typically relies on large suites of simulations to estimate covariances and construct emulators \citep{Kratochvil:2011eh, Armijo:2024ujo}. 
Forward-modelled field-level methods retain the complete spatial information within the data, offering the most direct route to unlocking the full potential of upcoming surveys \citep{Jasche_physical_2019, 2025JCAP...09..056S,2026arXiv260425385P}.

A key scientific target for upcoming surveys is to test the foundations of the cosmological model, particularly the nature of cosmic acceleration and the validity of General Relativity (GR) on large scales. While the $\Lambda$ Cold Dark Matter ($\Lambda$CDM) model successfully describes a wide range of observations, from the cosmic microwave background \citep{Planck_Planck_2020} to the expansion history of the Universe \citep{Riess_observational_1998} and the growth of structure \citep{Alam_completed_2021}, 
it relies on dark matter and dark energy, together comprising $\sim95\%$ of the Universe's energy density, whose physical nature remains unknown. Additionally, persistent tensions in key cosmological parameters such as the Hubble constant $H_0$ ($e.g.$, \citealt{Verde_tensions_2019,Riess_local_2024}) and hints of dynamic dark energy \citep{DESI_DR2_2025}, have further motivated the exploration of extensions to $\Lambda$CDM, including theories in which GR is modified on cosmological scales \citep{Koyama_cosmological_2016, Valentino_realm_2021, Abdalla_cosmology_2022}. Although GR has been rigorously tested in the strong-field and small-scale regimes through Solar System experiments \citep{Will_confrontation_2014, Mozaffari_tests_2015} and gravitational wave observations \citep{Baker_strong_2017}, its behaviour on cosmological scales remains comparatively less constrained \citep{Baker_novel_2019}.

Modified gravity (MG) theories often predict characteristic signatures in the growth and morphology of the LSS, particularly in the non-linear regime where screening mechanisms operate and departures from GR can become most pronounced \citep{Joyce_dark_2016, Baker_novel_2019}. The intricate network of clusters, filaments, walls and voids \citep{Bond_filaments_1996} encodes information in its morphology and evolution that is sensitive to the interplay between cosmic expansion, gravitational collapse and galaxy clustering, making it a powerful probe of gravitational physics. Because these signatures are inherently non-linear and environment-dependent, their detection is highly sensitive to the full field-level spatial arrangement of structure \citep{Coles_characterizing_2000, Leclercq_Bayesian_2015}. FLI is therefore a uniquely powerful approach for probing the nature of gravity, as it can extract information from the distribution of galaxies within cosmic filaments and under-dense voids and where departures from GR are typically the most pronounced \citep{Clampitt_voids_2013,Cai_testing_2015,Zivick_distinguishing_2016,2018MNRAS.475.3262F, Pisani_cosmic_2019}. Further, the environmental dependence of screening mechanisms is not captured by two-point statistics, but can be seen by the full field.

A central challenge in using the LSS to test gravity arises from the fact that galaxies do not perfectly trace the underlying dark matter distribution. The clustering of galaxies is related to the matter density field via galaxy bias relations that are inherently non-linear, scale-dependent and stochastic \citep{McDonald_clustering_2009,Desjacques_large_2018}. Crucially, the signatures of MG, such as the enhanced growth of structure driven by an environment-dependent fifth force, can be closely mimicked by changes in the galaxy bias. For instance, an observed increase in clustering power at a given scale may therefore reflect either departures from GR or a higher effective bias, leading to a strong degeneracy that limits the constraining power of two-point statistics. Traditional analyses are unable to disentangle these scenarios as they primarily encode the variance of the field rather than the morphological information that can differentiate gravitational dynamics from bias. In contrast, the Fourier phases capture the specific morphological configuration and environmental dependence of structure formation; these features are a direct consequence of gravitational physics and cannot be replicated by re-tuning the bias model. By forward-modelling the galaxy field itself, FLI allows MG parameters and galaxy bias to be constrained simultaneously; in this work, we demonstrate that it is specifically the inclusion of Fourier phase correlations within the FLI framework that predominantly breaks the inherent degeneracies between these parameters.

To jointly constrain MG and galaxy bias using the three-dimensional galaxy number counts field, we develop and apply a FLI pipeline. As a proof of concept, we focus on the Hu-Sawicki $f(R)$ model, a well-studied MG theory featuring a scalar degree of freedom and chameleon screening \citep{Khoury_chameleon_2004,Hu_models_2007}, and we couple this with a flexible non-linear galaxy bias prescription. Non-linear dark matter fields are generated across a range of gravity strengths using the COmoving Lagrangian Acceleration (COLA) method \citep{Tassev_solving_2013}. With our developed pipeline, we perform joint MG-bias parameter inference within a Bayesian likelihood framework: to assess the potential of an FLI analysis, we implement both a traditional power spectrum likelihood and a full field-level likelihood. We show that the FLI consistently yields tighter constraints than the power-spectrum-only analyses. 
We note that improvements over power-spectrum analyses are expected for any statistic that depends on higher-order $n$-point information. The information contained in the full field arises from both phase correlations and the non-Gaussian distribution of Fourier amplitudes. Our analysis quantifies the information gain arising from phase correlations and investigates FLI as an alternative means of capturing non-Gaussian information.

The remainder of this paper is organised as follows. In Section~\ref{sec:theory}, we outline the theoretical framework of this work, namely $f(R)$ gravity, the COLA method for non-linear structure formation, and our chosen galaxy bias prescription. Section~\ref{sec:method} details our methodology and simulation pipeline, including the simulation setup and generation of mock galaxy observables (Section~\ref{sec:method:setup}), the Bayesian inference frameworks for both the power spectrum and field-level analyses (Section~\ref{sec:method:likelihood}), and our parameter inference procedure (Section~\ref{sec:method:param_estimation}). Section~\ref{sec:Pofk_constraints} presents the results of the power spectrum analysis, covering shot noise estimation (Section~\ref{sec:Pofk_constraints:shot_noise}), the resulting MG-bias constraints  (Section~\ref{sec:Pofk_constraints:results}) and an exploration of the observed parameter degeneracies (Section~\ref{sec:Pofk_constraints:degeneracy}). These power spectrum results provide a baseline for our field-level analysis, which is presented in Section~\ref{sec:full_field_constraints:results}. We further investigate the contribution of different galaxy count voxels to the global likelihood (Section~\ref{sec:full_field_constraints:n_g_contributions}) and identify the cosmic environments driving the constraints (Section~\ref{sec:full_field_constraints:classifications}). In Section~\ref{sec:full_field_constraints:phase_only}, we explicitly isolate the role of Fourier phases in breaking the parameter degeneracies. The robustness of our inference pipeline is assessed in Section~\ref{sec:robustness} against variations in galaxy field thresholding (Section~\ref{sec:robustness:thresholding}), Poisson noise and initial conditions (Section~\ref{sec:robustness:Poisson_seeds}). Finally, we summarise and conclude in Section~\ref{sec:conclusions}.


\section{Theoretical framework}\label{sec:theory}

\subsection{Modified Gravity: $f(R)$ theories}\label{sec:theory:fofR}
The central aim of this work is to test whether field-level inference can disentangle the effects of Modified Gravity (MG) from galaxy bias on non-linear scales. As a proof of concept, we focus on the well-studied Hu-Sawicki $f(R)$ model \citep{Hu_models_2007}, which provides a representative example of scalar-tensor modifications to General Relativity (GR) and exhibits distinctive signatures in structure formation. This model is widely used in forecasts for Stage IV surveys such as DESI, Euclid and LSST, where gravity constraints increasingly rely on the information contained within the mildly to fully non-linear scales \citep{Amendola_cosmology_2018, Alam_towards_2021, 2025A&A...698A.233E}

In $f(R)$ gravity, the Einstein-Hilbert action is generalised by replacing the Ricci scalar, $R$, with a non-linear function, $f(R)$. The modified action is given by:
\begin{equation}
    S = \int d^4x \sqrt{-g}\left[\frac{1}{16\pi G} (R + f(R)) + \mathcal{L}_m \right],
\end{equation}
where $g$ is the determinant of the metric tensor, $G$ is the gravitational constant, and $\mathcal{L}_m$ is the Lagrangian of the matter fields. The function $f(R)$ introduces an additional scalar degree of freedom, which acts as a dynamical field and mediates a fifth force between matter. 

We consider the Hu-Sawicki formulation \citep{Hu_models_2007}, in which the model is characterised by a single free parameter, $f_{R0}$, controlling the strength of deviations from GR today:
\begin{equation}
    f(R) = -2 \Lambda + f_{R0}\frac{R_0^2}{R},
\end{equation}
where $\Lambda$ is a cosmological constant, $R_0$ is the present-day background Ricci scalar, and $-f_{R0}$ is the present-day scalar-field amplitude. Smaller values of $|f_{R0}|$ correspond to weaker modifications and GR is recovered in the limit $f_{R0}\rightarrow0$. Current constraints from astrophysical scales give $|f_{R0}| \lesssim 10^{-8}$ \citep{
Desmond:2020gzn, Landim:2024wzi} while a recent joint analysis of SPT clusters and Planck 2018 data yields $|f_{R0}| < 4.79 \times 10^{-6}$ \citep{SPT:2024adw}. Large scale structure gives slightly weaker constraints. A full shape analysis from BOSS DR12 gives $|f_{R0}| < 1.53 \times 10^{-5}$ \citep{Aviles:2024zlw} and a joint cosmic shear analysis combining DES-Y3, KiDS-1000, and HSC-Y3 together with CMB and BAO data yields $|f_{R0}| < 1.05 \times 10^{-5}$ \citep{Bai:2024hpw}.   

The additional fifth force enhances structure growth relative to GR, leading to observable signatures in the morphology of the cosmic web \citep{Falck:2015rsa}. To remain consistent with stringent Solar System tests of gravity \citep{Will_confrontation_2014,Mozaffari_tests_2015}, viable $f(R)$ models must possess a screening mechanism that suppresses this fifth force in high-density environments. In $f(R)$ gravity, screening is achieved via the Chameleon mechanism \citep{Khoury_chameleon_2004}, in which the scalar degree of freedom acquires a large effective mass in high-density regions, causing the fifth force to become short-ranged and decouple from matter in deep gravitational potentials. Consequently, the fifth force is suppressed in high-density regions and GR is locally recovered, while in low- and intermediate-density environments such as cosmic voids, there is a scale-dependent enhancement of growth of the linear density perturbations. This environment-dependent enhancement leads to characteristic signatures in the morphology of the cosmic web,
making $f(R)$ an ideal test case for assessing the ability of field-level inference to disentangle MG effects from galaxy bias. 

\subsection{Non-linear Structure Formation and COLA}\label{sec:theory:structure_formation}
A core requirement of field-level inference is the ability to forward-model the full three-dimensional matter field many times across the parameter space, generating the large ensembles of theoretical predictions needed for likelihood evaluation. 
This demands a method for modelling non-linear structure formation that is both accurate and computationally efficient. While full $N$-body simulations (see $e.g.$\ \citealt{Nbody_review}\ for a review) provide high accuracy, they are too computationally expensive for the thousands of realisations required in simulation-based inference. Within this work, we therefore adopt the COmoving Lagrangian Acceleration (COLA) method \citep{Tassev_solving_2013}, which provides a fast yet sufficiently accurate alternative, and for which extensions of this method to modified gravity theories are well-established and have been validated against full $N$-body solvers \citep{Winther_COLA_2017, 2021JCAP...09..021F, 2022JCAP...12..028F}. 

The LSS forms through the gravitational amplification of primordial density perturbations. On large scales and at early times, where these fluctuations are small ($\delta \equiv \delta\rho/\bar{\rho} \ll 1$), linear perturbation theory provides an accurate description and these fluctuations form a Gaussian random field \citep{Bardeen_statistics_1986}. As structure collapses under gravity, the evolution becomes non-linear, marking the transition towards the formation of the cosmic web, a complex network of halos, filaments, walls and voids \citep{Zeldovich_gravitational_1970,Bond_filaments_1996}. Lagrangian Perturbation Theory (LPT) provides a semi-analytic framework for describing these density perturbations in the mildly non-linear regime \citep{Buchert_Lagrangian_1993,Bernardeau_large-scale_2002}, while the deeply non-linear regime requires numerical solvers. 

COLA is specifically designed to capture both regimes efficiently. Large-scale displacements are computed analytically using second-order LPT (2LPT), while the remaining small-scale, non-linear dynamics are evolved numerically in a frame comoving with the 2LPT frame, using the particle-mesh (PM) techniques employed within full $N$-body codes. Because the large-scale bulk motion of the matter field is accounted for by 2LPT, only a modest number of timesteps is needed to evolve the residuals. This hybrid approach achieves a speed-up of $\mathcal{O}(10^2-10^3)$ relative to full $N$-body codes, while maintaining percent-level accuracy in the matter power spectrum up to $k\sim1 h$ Mpc$^{-1}$ \citep{Winther_COLA_2017, Izard_ICE-COLA_2018}. 

By combining the accuracy of 2LPT on large scales with a PM solver on small scales, COLA is an ideal tool for inference pipelines requiring large ensembles of theory predictions that require simulations. In this work we use an MG-extended implementation of COLA that includes $f(R)$ gravity, namely the Fourier-Multigrid Library\footnote{https://github.com/HAWinther/FML/} (\textsc{FML-COLA}). This implementation estimates the degree of screening by computing the gravitational potential, $\Phi_{N}$, of the density field and solving the linear field equation 
\citep{2015PhRvD..91l3507W, Winther_COLA_2017, 2021JCAP...09..021F}. While we utilise \textsc{FML-COLA}, our field-level framework is modular and could be adapted to use alternative gravitational forward models.

\subsection{Galaxy bias}\label{sec:theory:galaxy_bias}
One of the primary challenges in interpreting LSS data is modelling the relationship between the observed galaxy distribution and the underlying, invisible dark matter (DM) field. Galaxies preferentially form in high-density regions and are therefore biased tracers of the total matter distribution \citep{Kaiser_spatial_1984,Coles_bias_1999}. This bias is inherently complex, non-linear and scale-dependent ($e.g.$, \citealt{Fry_biasing_1993,Matsubara_diagrammatic_1995,Frusciante_Lagrangian_2012,Neyrinck_halo_2014,Desjacques_large_2018}). For tests of gravity, galaxy bias is a critical systematic: changes in the amplitudes of the power spectrum induced by MG, such as enhanced small-scale power from a fifth force, can be closely mimicked by changes in the bias parameters \citep{Devi_galaxy-halo_2019}, creating a degeneracy that limits the constraining power of two-point statistics. Any robust test of gravity using galaxy surveys must therefore treat the galaxy bias relation and gravitational parameters as part of a joint inference problem. 

In this work, we model the expected galaxy number count, $n_{\rm{g}}$, in each simulated voxel as a function of the underlying DM overdensity, $\delta_{\rm{DM}}$. Specifically, we adopt a four-parameter non-linear bias model proposed by \cite{Neyrinck_halo_2014} that captures both the enhancement of clustering in dense regions and the suppression of clustering in underdensities:
\begin{equation}\label{eq:bias}
    n_{\rm{g}} = N( 1 + \delta_{\rm{DM}} )^{\beta} \rm{exp}\left( -\rho( 1 + \delta_{\rm{DM}} )^{-\epsilon} \right) ,
\end{equation}
where $N$ sets the global normalisation, $\beta$ is the primary parameter controlling the non-linear response of galaxy clustering to local overdensities, and $\rho$ and $\epsilon$ describe the suppression of galaxy counts in low-density environments. The exponential term effectively downweights clustering in voids and other under-dense regions, while the term $(1 + \delta_{\rm{DM}})^{\beta}$ describes the non-linear clustering of galaxies. This model is one of many possible bias prescriptions that offer the flexibility required to jointly constrain gravitational dynamics and galaxy bias within our field-level framework. In this work, we use a local-in-density bias model, although recent work has shown that this cannot be fully realistic
\citep{Bartless_bye-bye_2024}. 

Eq.~\ref{eq:bias} is a phenomenological local-in-matter-density bias prescription rather than a systematically controlled perturbative bias expansion. Its form was motivated by N-body measurements of the halo–matter relation, particularly the power-law behaviour at moderate density and exponential suppression in voids \citep{Neyrinck_halo_2014}. Closely related implementations have been used in BORG field-level analyses of the 2M++ galaxy catalogue at a resolution of 2.65 Mpc/h \citep{Jasche_physical_2019, Stopyra:2023yqm} providing practical precedent at a spatial resolution comparable to the 3.125 Mpc/h voxels used here. This precedent should not, however, be interpreted as establishing the accuracy of the model on few-Mpc scales. Indeed, controlled halo-field tests find increasing sensitivity to bias-model assumptions towards finer grid resolution~\citep{Nguyen_2021}. We therefore employ Eq.~\ref{eq:bias} only as a controlled phenomenological model for the proof-of-concept experiment below; a real-data application on these scales would require a more complete treatment of local and non-local bias and stochasticity.


\section{Methodology}\label{sec:method}
The central motivation of this work is to test whether field-level inference (FLI) can provide enhanced constraints on MG as compared to traditional analyses using the power spectrum. Because we observe galaxies rather than dark matter, we must forward-model the full galaxy distribution. By preserving Fourier phase information and capturing higher-order non-Gaussian features, this galaxy field retains the information necessary to break degeneracies between MG and galaxy bias that is otherwise lost when compressing the field into two-point statistics. 

We develop a simulation-based pipeline that enables a direct, field-level comparison between the predicted galaxy number counts field in MG and GR models. This pipeline consists of three primary modules: (i) an ensemble of fast dark matter simulations combined with a non-linear galaxy bias model to generate mock galaxy observables; (ii) Bayesian likelihood frameworks for both the power spectrum and field-level analyses; (iii) parameter inference via the interpolation of a pre-computed likelihood grid. This modular design allows us to contrast the constraining power of traditional $P_{\rm{g}}(k)$ analyses with that of the full-field approach under controlled, repeatable conditions. 

In this section, we present our simulation and inference pipeline. We first describe the setup of our COLA simulations and the generation of mock galaxy observables, $i.e.$, the galaxy number counts field, $n_{\rm{g}}$, and the galaxy power spectrum, $P_{\rm{g}}(k)$ (Section~\ref{sec:method:setup}). We then detail the statistical frameworks used to infer model parameters in Section~\ref{sec:method:likelihood}; we assume a Gaussian likelihood for the power-spectrum-only analysis, whereas the field-level inference employs both a full-field Poisson likelihood and a truncated Poisson likelihood designed for thresholded data. Finally, Section~\ref{sec:method:param_estimation} outlines the parameter inference strategy based on interpolating a pre-computed likelihood grid to explore the joint MG and galaxy bias parameter space.

\subsection{Simulation setup and mock generation}\label{sec:method:setup}
Our simulation-based pipeline requires two galaxy observables: the number counts field, $n_{\rm{g}}$, and the power spectrum, $P_{\rm{g}}(k)$. In this section, we describe how we generate the mock galaxy observables and the predicted fields used for parameter inference. 

To perform Bayesian inference, we sample the joint parameter space of galaxy bias and modified gravity, $\theta = (\beta, |f_{R0}|)$. As an initial demonstration of our pipeline, we focus on a simplified case where only the primary bias parameter, $\beta$, is varied: all predicted $n_{\rm{g}}$ fields (and their corresponding $P_{\rm{g}}(k)$) are computed by fixing the suppression parameters $\rho$ and $\epsilon$ (Eq.~\ref{eq:bias}) to their fiducial values. These parameters primarily control the suppression of galaxy formation in low-density regions and are sensitive to the simulation resolution.Extending our analysis to include $\rho$ and $\epsilon$ within the parameter space, as well as extending to more complex galaxy bias models such as those including non-locality and stochasticity, is reserved for future work. Our core objective here is to conduct a joint analysis between modified gravity ($|f_{R0}|$) and the overall magnitude of galaxy bias ($\beta$).

Additionally, the normalisation $N$ is tuned according to the  $(\beta, |f_{R0}|)$ values: 
\begin{equation}
N = N(\beta, |f_{R0}|),
\end{equation}
to ensure that the total number of galaxies, $N_{\rm{tot}}$, remains constant across all gravity models and all bias models considered. Importantly, this tuning procedure guarantees that differences in the likelihood are due to the subtle differences in the spatial distribution of galaxies across models, rather than due to any difference in $N_{\rm{tot}}$: in fact, while the galaxy power spectrum is not dependent on $N_{\rm{tot}}$, the full field itself is extremely sensitive to changes in the total galaxy number count and this sensitivity is reflected in the computed likelihoods of test models. 
We distinguish between three distinct $n_g$ predictions:
\begin{enumerate}
    \item The fiducial truth, $n_{\rm{g}}^{\rm{fid}}$, which is the noiseless theoretical prediction for the fiducial model parameters, $\theta^{\rm{fid}} = (\beta^{\rm{fid}},|f_{R0}|^{\rm{fid}})$. 
    \item The observed mock, $n_{\rm{g}}^{\rm{obs}}$, which is the noisy data field generated by Poisson-sampling $n_{\rm{g}}^{\rm{fid}}$.
    \item The model prediction, $n_{\rm{g}}^{\rm{pred}}(\theta)$, which is the noiseless theoretical prediction generated for any test parameter set $\theta = (\beta, |f_{R0}|)$.
\end{enumerate}
We denote the corresponding power spectra of these galaxy number counts fields as $P_{\rm{g}}^{\rm{fid}}(k)$, $P_{\rm{g}}^{\rm{obs}}(k)$ and $P_{\rm{g}}^{\rm{pred}}(k)$, respectively. 

To compute both $n_{\rm{g}}$ and $P_{\rm{g}}(k)$, we start from the underlying dark matter mock. These mocks are generated for specific gravity models using the \textsc{FML-COLA} code, an implementation of COLA for running modified gravity simulations (see Section~\ref{sec:theory:structure_formation}). All simulations presented in this work evolve $512^3$ particles from redshift $z=19$ to $z=0$ within a cubic volume of side length $400\ h^{-1}$ Mpc, and forces are evaluated on a $1536^3$ grid. We assume a fixed $\Lambda$CDM cosmology consistent with Planck \citep{Planck_Planck_2014}: $\Omega_m=0.307,\ \Omega_{\Lambda}=0.693,\ \Omega_b=0.04825,\ h=0.705,\ \sigma_8=0.8288,\ n_s=0.9611$. We follow a similar specification to that presented in \cite{Lilow_neural_2024}.

Gaussian initial conditions are generated at $z=19$ using 2LPT. From the final COLA snapshot at $z=0$, we use the \textsc{Pylians} library\footnote{https://github.com/franciscovillaescusa/Pylians} \citep{Pylians} to deposit the particles onto a regular $128^3$ grid via Cloud-in-Cell (CIC) mass-assignment, producing the dark matter overdensity field, $\delta_{\rm{DM}}$. 

The mean expected galaxy number counts field, $n_{\rm{g}}$, is then computed by applying the non-linear bias model presented in Eq.~\ref{eq:bias} to $\delta_{\rm{DM}}$. We consider two fiducial scenarios:  
\begin{enumerate}
    \item \textbf{GR}: $|f_{R0}|=0$,
    \item \textbf{F6}: $|f_{R0}|=10^{-6}$.
\end{enumerate}
This allows us to assess the constraining power of our FLI approach in the case of both GR and MG, for which we choose a relatively small modification to gravity. We generate the fiducial DM overdensity field, $\delta_{\rm{DM}}^{\rm{fid}}$, for each case, and then compute $n_{\rm{g}}^{\rm{fid}}$ by applying Eq.~\ref{eq:bias} with bias parameters $N=0.19$, $\beta=0.77$, $\rho=1.61$ and $\epsilon=0.09$. The values adopted here are the ensemble-mean values inferred for sample 11 of the real-data 2M++ BORG analysis of \cite{Jasche_physical_2019}, in which all bias parameters were instead marginalised jointly with the density field. We fix $\rho$ and $\epsilon$ deliberately to reduce the present experiment to the two-dimensional ($\beta, f_{R0}$) problem. Consequently, the constraints obtained below are conditional on the assumed values and should be regarded as optimistic relative to a real-data analysis in which these and additional bias degrees of freedom must be marginalised.

With our simulation setup, this yields a total of $N_{\rm{tot}}\sim 69,000$ galaxies for both fiducial models considered, roughly matching the survey size of the 2M++ catalogue \citep{Lavaux_2M++_2011}.

From $n_{\rm{g}}^{\rm{fid}}$, we generate the noisy observational mock, $n_{\rm{g}}^{\rm{obs}}$, by sampling each voxel, $i$, from a Poisson distribution:
\begin{equation}
n_{\rm{g},i}^{\rm{obs}} \sim \rm{Poisson}(n_{\rm{g},i}^{\rm{fid}}),
\end{equation}
where $n_{\rm{g},i}^{\rm{fid}}$ is the fiducial galaxy count in voxel $i$ and $n_{\rm{g},i}^{\rm{obs}}$ is the resulting discrete number count, including Poisson noise. 

The galaxy power spectra are computed using the \textsc{Pylians}\footnote{https://pylians3.readthedocs.io/en/master/index.html} library. For any galaxy number counts field $n_{\rm{g}}$ (whether fiducial, observational or predicted), the galaxy overdensity field, $\delta_{\rm{g}}$, is constructed as:
\begin{equation}\label{eq:delta_g}
    \delta_{\rm{g}} = n_{\rm{g}} / \langle n_{\rm{g}}\rangle - 1,   
\end{equation}
where $\langle \rangle$ denotes the volume average. The overdensity field is Fast Fourier Transformed and the three-dimensional power spectrum is computed and spherically averaged by binning Fourier modes in shells of constant $k = |\mathbf{k}|$. The Nyquist frequency is $k_{\rm{Nyq}}=\pi N_{\rm{grid}}/L_{\rm{box}}$. To avoid resolution effects, we restrict the analysis to modes where $k < k_{\rm{Nyq}}/2$. 

For parameter inference, we generate $n_{\rm{g}}^{\rm{pred}}(\theta)$ and the corresponding $P_{\rm{g}}^{\rm{pred}}(\theta, k)$ for each test model in our $(\beta, |f_{R0}|)$ parameter space. The aforementioned renormalisation of $N(\beta, |f_{R0}|)$ is handled at the time of run, so that the predicted fields are computed as $n_{\rm{g}}^{\rm{pred}}(\beta,N(\beta, |f_{R0}|),\rho_{\rm{fid}},\epsilon_{\rm{fid}})$. This enables us to efficiently obtain joint constraints on $\beta$ and $|f_{R0}|$. These theoretical predictions remain noiseless; we do not add any Poisson to the fields of the test models. Using noiseless predictions allows us to directly compare the discrete noisy data, $n_{\rm{g}}^{\rm{obs}}$, to the expectation values, $n_{\rm{g}}^{\rm{pred}}(\theta)$, by evaluating the Poisson likelihood within each voxel.

To ensure that any differences in our results are driven solely by the physical parameters, our pipeline is built to ensure that all fields ($i.e.$, both $n_{\rm{g}}$ and $P_{\rm{g}}(k)$ of the fiducial truth, the observational mocks and the model predictions) are generated using the exact same realisation of Gaussian initial conditions and the same random seed for Poisson noise.

\begin{figure*}
    \centering
    \includegraphics[width=\textwidth]{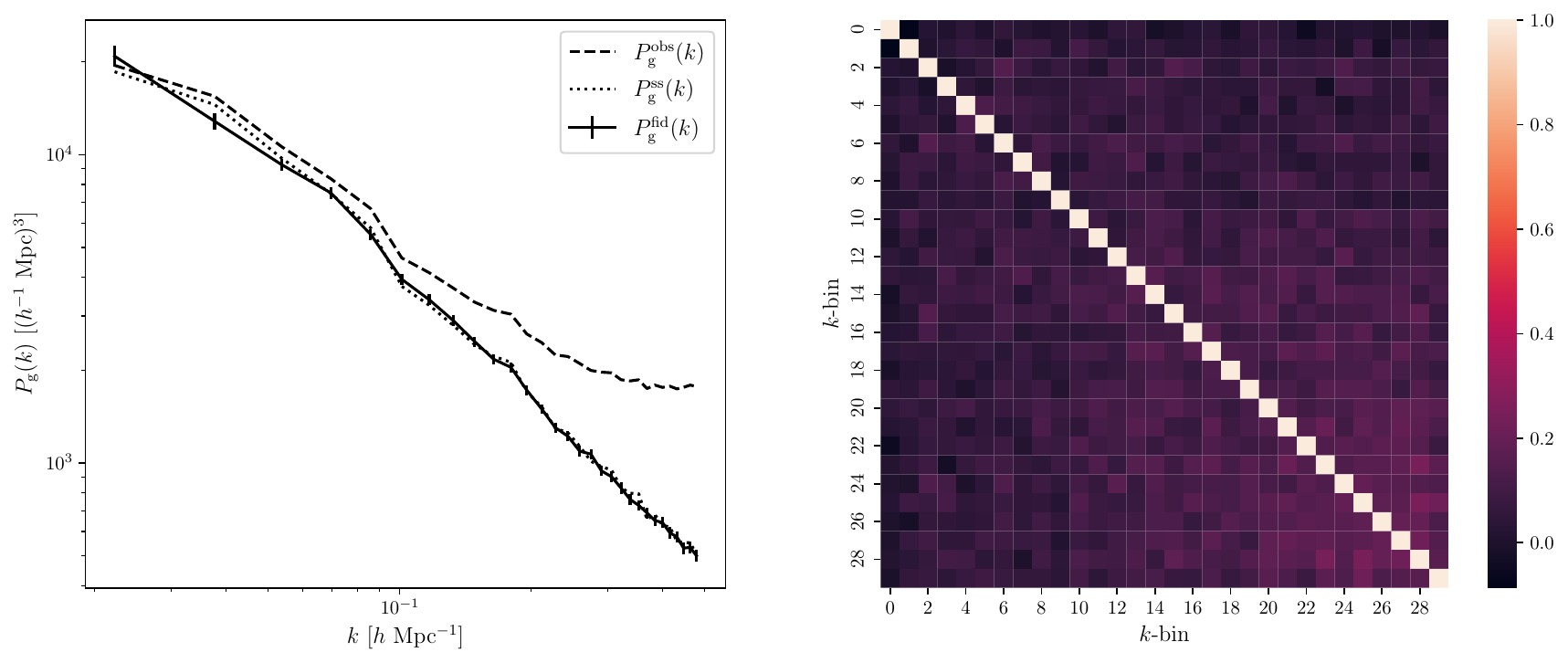}
    \caption{\textbf{Left:} The galaxy power spectrum, $P_{\rm{g}}(k)$, for the GR fiducial model. We show the noiseless theoretical prediction, $P_{\rm{g}}^{\rm{fid}}(k)$, with $1\sigma$ uncertainties estimated from the diagonal of the covariance matrix, $\mathbf{C}$, of $N_{\rm{mock}}=1024$ independent Poisson-sampled realisations (solid line with error bars). 
    We compare this against the power spectrum measured from the noisy mock data field, $P_{\rm{g}}^{\rm{obs}}(k)$ (dashed line), and the corresponding shot-subtracted power spectrum, $P_{\rm{g}}^{\rm{ss}}(k)$ (dotted line). After subtracting the shot noise contribution, estimated using Eq.~\ref{eq:shot_noise}, the recovered $P_{\rm{g}}^{\rm{ss}}(k)$ agrees with the fiducial prediction, $P_{\rm{g}}^{\rm{fid}}(k)$, within cosmic variance. \textbf{Right:} The correlation matrix of the shot-subtracted galaxy power spectra, $P_{\rm{g}}^{\rm{ss}}(k)$, estimated from $N_{\rm{mock}}=1024$ realisations, highlighting mode correlations at smaller scales.
  }
    \label{fig:Pofk_shot_noise_demonstration}
\end{figure*}

\subsection{Bayesian Inference Framework}\label{sec:method:likelihood}
To obtain constraints on our model parameters, $\theta = (\beta, |f_{R0}|)$, we adopt a Bayesian framework. In this work, we compare the traditional approach of using the power spectrum against our new field-level inference method. We first define the Gaussian likelihood for the power spectrum in Section~\ref{sec:method:likelihoods:Gaussian}, which serves as our baseline for parameter constraints. We then present the likelihoods for the field-level analysis, for which we consider two cases: a full-field Poisson likelihood (Section ~\ref{sec:method:likelihoods:Poisson}) to capture the total information content of the galaxy distribution, and a truncated Poisson likelihood (Section ~\ref{sec:method:likelihoods:truncPoisson}) to investigate the robustness of our method against data thresholding, representing scenarios where low-density information might be excluded in observational data. In all cases, the posterior probability distribution for a given data vector $\textbf{D}$ (representing either the field $n_{\rm{g}}^{\rm{obs}}$ or the power spectrum $\mathbf{P}_{\rm{g}}^{\rm{obs}}$), $P(\theta \mid \mathbf{D})$, follows Bayes' theorem:
\begin{equation}
    P(\theta \mid \mathbf{D}) \propto \mathcal{L}(\mathbf{D} \mid \theta) P(\theta)
\end{equation}
where $\mathcal{L}(\mathbf{D} \mid \theta)$ is the likelihood associated with the chosen statistics and $P(\theta)$ is the prior.

\subsubsection{Power Spectrum Gaussian Likelihood}\label{sec:method:likelihoods:Gaussian}
To establish a baseline for comparison with our field-level inference pipeline, we implement a traditional power spectrum analysis. The galaxy power spectrum, $P_{\rm{g}}(k)$, quantifies the clustering amplitude at a given scale, represented by the wavenumber $k$. The power spectrum captures the amplitudes of the Fourier modes, but disregards some of the non-Gaussian information such as non-Gaussian phase correlations.

We assume a Gaussian likelihood function of the form:
\begin{equation}\label{eq:Gaussian_likelihood}
    \ln \mathcal{L}(\mathbf{P}^{\rm{obs}} \mid \theta) = -\frac{1}{2} \left[\mathbf{P}^{\rm{obs}} - \mathbf{P}^{\rm{pred}}(\theta) \right]^T \mathbf{C}^{-1}_{\rm{corr}} \left[ \mathbf{P}^{\rm{obs}} - \mathbf{P}^{\rm{pred}}(\theta) \right],
\end{equation}
where $\mathbf{P}^{\rm{obs}}$ is the mock data vector at the sampled $\mathbf{k}$ bins, $\mathbf{P}^{\rm{pred}}(\theta)$ is the theoretical prediction vector at the same $\mathbf{k}$, and $\mathbf{C}^{-1}_{\rm{corr}} = \mathcal{F} \mathbf{C}^{-1}$ is the corrected inverse covariance matrix. We estimate $\mathbf{C}$ from the power spectra of 1024 independent Poisson realisations of the noiseless fiducial $n_{\rm{g}}^{\rm{fid}}$ field, and we apply the Percival correction factor, $\mathcal{F}$ (Eqs. 55 and 56 in \citealt{Percival_matching_2022}), to its inverse to account for the underestimation of errors introduced by using a finite number of realisations. The power spectrum covariance excludes sample variance, ensuring a consistent comparison with the fixed-phase field-level inference results.

\subsubsection{Full-field Poisson Likelihood}\label{sec:method:likelihoods:Poisson}
Our primary analysis is a field-level approach that computes the Poisson likelihood $\mathcal{L}(n_{\rm{g}}^{\rm{obs}}\mid\theta)$ on a voxel-by-voxel basis of the full 3D galaxy number counts field. The number of galaxies in each voxel $i$ is expected to follow an independent Poisson distribution with predicted mean count $\lambda_i(\theta)
=n_{\rm{g},i}^{\rm{pred}}(\theta)$, given the parameters $\theta=(\beta, |f_{R0}|)$. 
The probability of observing $n_{\rm{g},i}^{\rm{obs}}$ galaxies in voxel $i$ is therefore
\begin{equation}
    P(n_{\rm{g},i}^{\rm{obs}} \mid \theta) = \frac{\lambda_i(\theta)^{\,n_{\rm{g},i}^{\rm{obs}}} \, e^{-\lambda_i(\theta)}}{n_{\rm{g},i}^{\rm{obs}}!}.
\end{equation}

The corresponding voxel-level log-likelihood is
\begin{equation}
    \ln \mathcal{L}_i(n_{\rm{g},i}^{\rm{obs}} \mid \theta) = n_{\rm{g},i}^{\rm{obs}} \ln \lambda_i(\theta) - \lambda_i(\theta) - \ln(n_{\rm{g},i}^{\rm{obs}}!),
\end{equation}
where the final term acts as a constant offset throughout the parameter grid due to its independence of $\theta$. The global log-likelihood of the full field is then obtained by summing over all $N_{\rm{vox}}$ voxels:
\begin{equation}\label{eq:Poisson_likelihood}
    \ln \mathcal{L}(n_{\rm{g}}^{\rm{obs}} \mid \theta) = \sum_{i=1}^{N_{\rm{vox}}} \ln \mathcal{L}_i(n_{\rm{g},i}^{\rm{obs}}|\theta).
\end{equation}

Unlike the Gaussian likelihood of the power spectrum, this approach preserves both Fourier amplitudes and phases, thus retaining the full statistical information of the field. In addition to storing the summed (global) likelihood, $\mathcal{L}(n_{\rm{g}}^{\rm{obs}} \mid \theta)$, we save the full 3D box of voxel-level likelihood contributions, $\mathcal{L}_i$, enabling us to investigate which cosmic environments drive the parameter constraints (see Section~\ref{sec:full_field_constraints:classifications}). 

\begin{figure*}
    \centering
    \includegraphics[width=\textwidth]{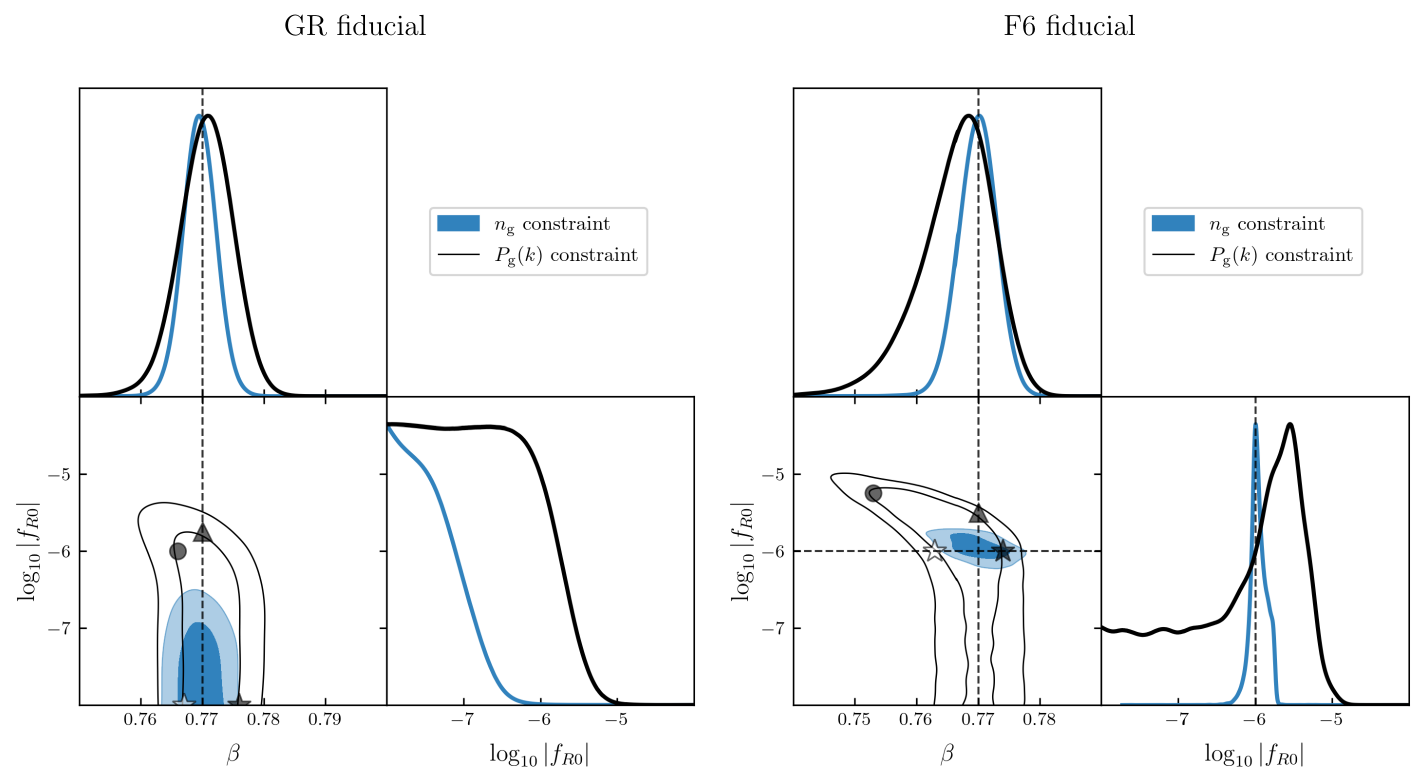} 
    \caption{
    Comparison of constraints on $\beta$ and $\log_{10}|f_{R0}|$ obtained with the power spectra of the galaxy number count, $P_{\rm{g}}(k)$, and the full-field analysis, $n_{\rm g}$. The former uses the Gaussian likelihood in Eq.~\ref{eq:Gaussian_likelihood}, whereas the latter employs the Poisson likelihood in Eq.~\ref{eq:Poisson_likelihood}. \textbf{Left:} GR fiducial mock (true values: $\log_{10}|f_{R0}|=-\infty, \beta=0.77$; prior ranges: $\beta \in [0.75,0.80]$, $\log_{10}|f_{R0}| \in [-8,-4]$); \textbf{Right:} F6 fiducial mock (true values: $\log_{10}|f_{R0}|=-6, \beta=0.77$; prior ranges: $\beta \in [0.74,0.79]$, $\log_{10}|f_{R0}| \in [-8,-4]$). The contours show the 1$\sigma$ and 2$\sigma$ confidence intervals. Dashed lines indicate the true fiducial values (note that the true $\log_{10}|f_{R0}|$ value for the GR fiducial case lies beyond our prior range and is therefore not shown). 
    The power-spectrum constraints show a degeneracy in ($\beta$, $\log_{10}|f_{R0}|$) that we analyse further in Figure~\ref{fig:Pofk_2sigma}, where we show the power spectra of models lying at the $1\sigma$ posterior boundaries. The investigated $(\beta,|f_{R0}|)$ combinations are marked on the contours here for comparison: filled and hollow stars represent models with increasing and decreasing galaxy bias, respectively; filled triangles represent models with increased gravity strength; finally, filled circles represent a model along the degeneracy line.
    }  
    \label{fig:n_g_constraint}
\end{figure*}
 
\begin{table*}
\centering
\caption{
Comparison of marginalised constraints from the $P_{\rm{g}}(k)$ and $n_{\rm{g}}$ (full-field) analyses for both fiducial models considered (GR and F6). We state the 68\% confidence intervals for $\beta$, and state either the 95\% upper limit (GR fiducial case) or the 68\% confidence intervals (F6 fiducial case) for $\log_{10}|f_{R0}|$. The true values are as follows: GR fiducial model has $\log_{10}|f_{R0}|=-\infty$ and $\beta=0.77$; F6 fiducial model has $\log_{10}|f_{R0}|=-6$ and $\beta=0.77$. Note the considerable tightening of constraints in the $n_{\rm{g}}$ analysis. }
\renewcommand{\arraystretch}{1.3}
\setlength{\tabcolsep}{10pt}
\begin{tabular}{||c||c|c||c|c||}
\hline
\multirow{2}{*}{Analysis} & \multicolumn{2}{c||}{\textbf{GR Fiducial}} & \multicolumn{2}{c||}{\textbf{F6 Fiducial}} \\
\cline{2-5}
 & $\beta$ & $\log_{10}|f_{R0}|$ & $\beta$ & $\log_{10}|f_{R0}|$ \\
\hline\hline
$P_{\rm{g}}(k)$ & $0.771^{+0.005}_{-0.004}$ & $< -5.698$ & $0.766^{+0.007}_{-0.004}$ & $-6.223^{+1.032}_{-0.411}$ \\
\hline
$n_{\rm{g}}$  & $0.770\pm0.003$ & $< -6.760$ & $0.770\pm0.003$ & $-5.958^{+0.100}_{-0.107}$ \\
\hline
\end{tabular}
\label{tab:Pofk_vs_n_g_summary}
\end{table*}

\subsubsection{Truncated Poisson Likelihood}\label{sec:method:likelihoods:truncPoisson}
To assess the robustness of our likelihood inference pipeline, we also consider cases where a threshold is applied to $n_{\rm{g}}^{\rm{obs}}$ (see Section~\ref{sec:robustness:thresholding}). We implement this approach to account for the inherent uncertainties in observational surveys, such as magnitude-limited detections whereby voxels with low number counts may be excluded. Within our pipeline, voxels within $n_{\rm{g}}^{\rm{obs}}$ that have a galaxy number count less than a chosen threshold $T$ ($i.e.$, $n_{\rm{g,i}}^{\rm{obs}}<T$), are excluded from the likelihood evaluation. These voxels are identified according to $n_{\rm{g}}^{\rm{obs}}$, but are excluded from both $n_{\rm{g}}^{\rm{obs}}$ and $n_{\rm{g}}^{\rm{pred}}(\theta)$ before performing the inference. 

In this scenario, we assume the likelihood follows the form of a truncated Poisson distribution, in which the standard Poisson probability mass function must be re-normalised by the probability of observing a count at or above the threshold, in order to account for the truncated distribution. For a voxel $i$ with predicted mean count $\lambda_i(\theta)=n_{\rm{g},i}^{\rm{pred}}(\theta)$ and observed count $n_{\rm{g},i}^{\rm{obs}}$, we define the normalisation factor for truncation, $\mathcal{Z}_i(\theta,T)$, as follows:

\begin{equation}
    \mathcal{Z}_i(\theta,T) = 1 - \sum_{m=0}^{T-1} \dfrac{\lambda_i(\theta)^{\,m} e^{-\lambda_i(\theta)}}{m!}
    \label{eq:trunc_poisson_norm}
\end{equation}
so that the truncated Poisson likelihood of observing $n_{\rm{g},i}^{\rm{obs}}$ galaxies in voxel $i$, conditional on $n_{\rm{g},i}^{\rm{obs}}\geq T$, takes the form:
\begin{equation}
    P(n_{\rm g,i}^{\rm{obs}} \mid \theta,\, n_{\rm g,i}^{\rm{obs}} \geq T) 
    = \frac{P(n_{\rm g,i}^{\rm{obs}} \mid \theta)}{\mathcal{Z}_i(\theta, T)}.
    \label{eq:trunc_poisson}
\end{equation}

The denominator ensures that probabilities are properly normalised once the voxels $n_{\rm{g},i}^{\rm{obs}}<T$ are discarded. Here, $P(n_{\rm g,i}^{\rm{obs}} \mid \theta)$ is the full-field Poisson likelihood defined in Section~\ref{sec:method:likelihoods:Poisson} and $m$ is a summation index that runs over galaxy number counts below the truncation threshold.

The voxel-level truncated log-likelihood is then:
\begin{equation}
\begin{split}
    \ln \mathcal{L}_i(n_{\rm{g},i}^{\rm{obs}}\mid\theta, n_{\rm{g},i}^{\rm{obs}} \geq T) = n_{\rm{g},i}^{\rm{obs}} \ln \lambda_i(\theta) - \lambda_i(\theta) - \ln(n_{\rm{g},i}^{\rm{obs}}!) \\
     - \ln \left( \mathcal{Z}_i(\theta, T) \right),
\end{split}
\end{equation}
where the last term accounts for the renormalisation after thresholding. By summing over the voxels that remain after satisfying the thresholding condition, we obtain the global log-likelihood:
\begin{equation}\label{eq:truncPoisson_likelihood}
    \ln \mathcal{L}(n_{\rm{g}}^{\rm{obs}} \mid \theta, n_{\rm{g}}^{\rm{obs}} \geq T) = \sum_{i=1}^{N_{\rm{vox}}} \ln \mathcal{L}_i(n_{\rm{g},i}^{\rm{obs}}|\theta, n_{\rm{g},i}^{\rm{obs}} \geq T).
\end{equation}

This truncated Poisson likelihood is required to avoid biased estimates of the galaxy bias parameter $\beta$, which would otherwise arise if only high-density regions were selected without accounting for the change in the probability normalisation. We note that this is an approximate method to account for selection effects; a more principled formalism within a Bayesian inference framework is presented in \cite{Kelly:2008wk, Desmond:2025ggt, Stiskalek_1.8_2026}.

\subsection{Parameter Inference}\label{sec:method:param_estimation}
We generate a 2D grid of theoretical predictions of the galaxy number counts field, $n_{\rm{g}}^{\rm{pred}}(\beta, |f_{R0}|)$, following the procedure detailed in Section~\ref{sec:method:setup}. To explore the $|f_{R0}|$ parameter space, we run a total of 18 COLA simulations: $|f_{R0}|=0$ (GR) and 17 values spaced logarithmically from $10^{-8}$ to $10^{-4}$ ($\Delta\log_{10}|f_{R0}|=0.25$). For each gravity model, we generate 55 $n_{\rm{g}}^{\rm{pred}}(\theta)$ fields and their corresponding power spectra $P_{\rm{g}}^{\rm{pred}}(\theta)$, varying $\beta$ by linearly separated intervals within a defined prior range. This gives a total of $18 \times 55 = 990$ model predictions for each likelihood evaluation.


We compute the likelihood, $\mathcal{L}(\mathbf{D} \mid \theta)$, for each of the 990 discrete $(\beta,|f_{R0}|)$ combinations using the appropriate framework: Eq.~\ref{eq:Gaussian_likelihood} for $P_{\rm{g}}(k)$ analysis, Eq.~\ref{eq:Poisson_likelihood} for the full field-level inference, or Eq.~\ref{eq:trunc_poisson} for a thresholded field. Since Markov Chain Monte Carlo (MCMC) methods require a continuous likelihood function surface, we interpolate this discrete grid of likelihood values using a linear regular grid interpolator. We verified that the interpolated likelihood is smooth throughout the region of parameter space explored by the MCMC chains by analysing the likelihood values sampled directly from the chains. We found no evidence of interpolation-induced discontinuities.

This continuous function is then passed to the MCMC sampler \textsc{emcee}\footnote{https://github.com/dfm/emcee} \citep{emcee} to explore the parameter space and obtain a posterior distribution, $P(\theta\mid\mathbf{D})$. For all MCMC analyses, we utilise 20 walkers and ensure convergence by imposing the following criteria: (i) the total chain length exceeds $100$ times the integrated autocorrelation time, $\tau$, for all parameters (a total of two parameters in this case), $i.e.$, $N_{\rm{iter}} > 100\ \tau$; (ii) the relative change in $\tau$ between successive estimates (computed every $100$ iterations) satisfies $\left|\tau_{\rm{old}} - \tau_{\rm{new}}\right|/\tau_{\rm{new}} < 0.01$. We note that \textsc{emcee} only samples within the prior range for $f_{R0}$ of $\log_{10}|f_{R0}| \in [-8,-4]$, as it is not possible to sample in log-space up to the true $f_{R0}$ value in GR ($\log_{10}|f_{R0}| = -\infty$). The prior range of $\beta$ is selected to be sufficiently broad that the posterior is not affected by the prior boundaries. 

We also note that the constraint on $\log_{10}|f_{R0}|$ depends on the lower prior bound. To obtain a prior-independent bound, one could follow the approach proposed in \cite{Gordon_bayesian_2007}. However, as our primary aim is to contrast the constraining power of the galaxy power spectrum and the full galaxy number counts field, both of which use the same prior, we do not implement this correction in this paper.


\begin{figure*}
    \centering
    \includegraphics[width=\linewidth]{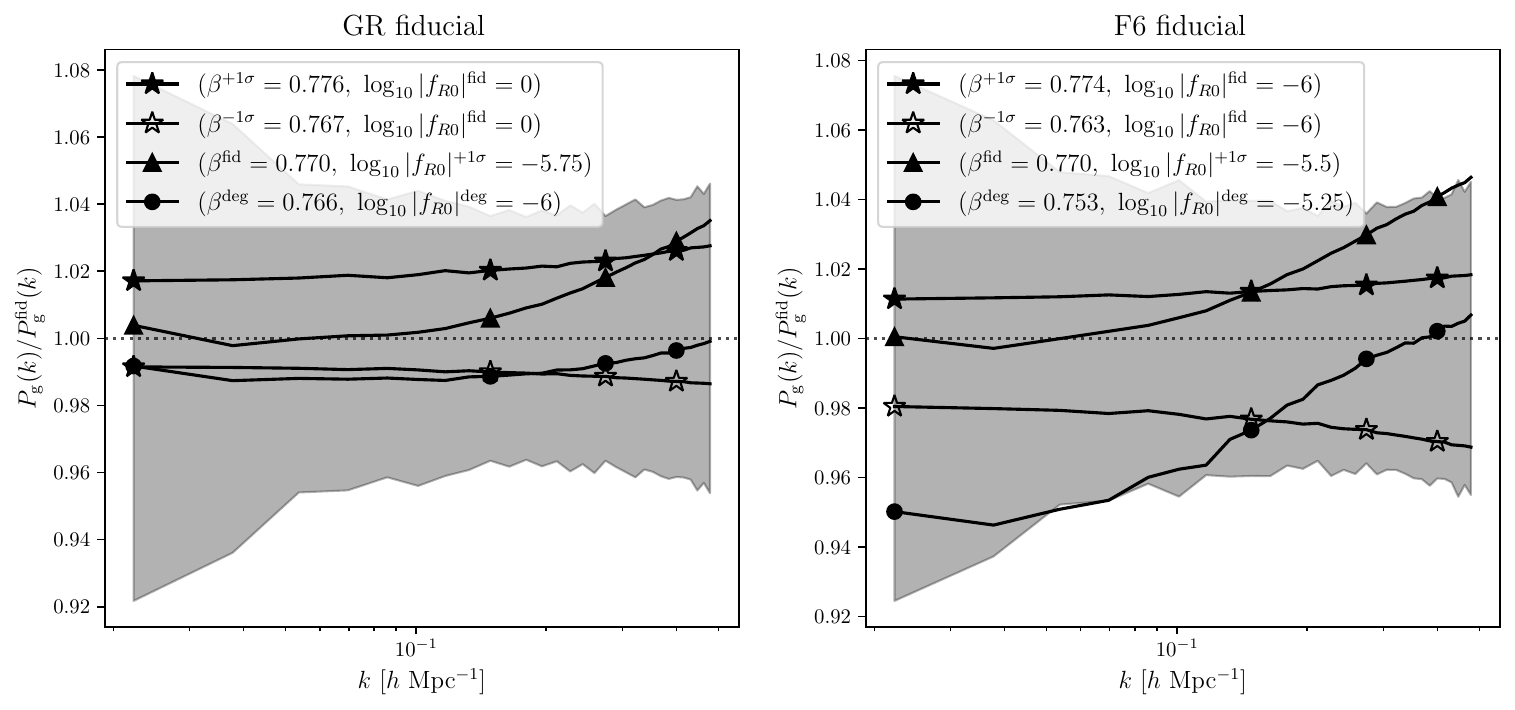}
    \caption{Ratio of the predicted galaxy power spectra, $P_{\rm{g}}(k)$, relative to the fiducial $P_{\rm{g}}^{\rm{fid}}(k)$, for models indicative of increasing/decreasing galaxy bias (filled/hollow stars, respectively), increasing gravity strength (triangles), and a degeneracy-line model (circles). The specific marker shapes and fills used in this figure exactly match the corresponding models shown in Figure~\ref{fig:n_g_constraint}. Combinations of $(\beta,|f_{R0}|)$ are chosen according to the $1\sigma$ bounds of the $P_{\rm{g}}(k)$ analysis shown in Figure~\ref{fig:n_g_constraint}, with the superscripts "$\pm1\sigma$", "$\rm{fid}$" and "$\rm{deg}$" denoting values of the $\pm1\sigma$ boundaries, the fiducial model and the degeneracy-line model, respectively. \textbf{Left:} GR fiducial. \textbf{Right:} F6 fiducial. The grey shaded region represents the $1\sigma$ boundary computed from the diagonal of the covariance matrix estimated with $N_{\rm{mock}}=1024$ independent Poisson-sampled realisations of $n_{\rm{g}}^{\rm{fid}}$. Each of these models, representing changes in modified gravity or galaxy bias, is indistinguishable within the statistical uncertainties when considering power spectrum information alone. }
    \label{fig:Pofk_2sigma}
\end{figure*}

\section{Power Spectrum Analysis}\label{sec:Pofk_constraints}
We begin by establishing baseline constraints using the two-point statistic of the galaxy field. The galaxy power spectrum, $P_{\rm{g}}(k)$, is the traditional summary statistic of large-scale structure studies and has been used to obtain constraints on modified gravity ($e.g.$, \citealt{Dossett_constraining_2014, Moretti_modified_2023,Piga_constraints_2023, Rodriguez-Meza_fkPT_2024}). 

In this section, we present our $P_{\rm{g}}(k)$ analysis and recovered constraints on $(\beta,|f_{R0}|)$. All results are derived from a single realisation of Gaussian initial conditions used to generate $\delta_{\rm{DM}}$ with COLA, and a single realisation of Poisson noise added to the noiseless fiducial field, $n_{\rm{g}}^{\rm{fid}}$, from which we generate the mock observed field $n_{\rm{g}}^{\rm{obs}}$ and the corresponding observed power spectrum, $P_{\rm{g}}^{\rm{obs}}(k)$ (see Section~\ref{sec:method:setup} for details). 

We first describe our procedure for estimating and subtracting the shot noise contribution from the mock galaxy power spectra (Section~\ref{sec:Pofk_constraints:shot_noise}). In Section~\ref{sec:Pofk_constraints:results}, we present the posterior distributions for our two fiducial scenarios considered (GR and $|f_{R0}|=10^{-6}$; see Section~\ref{sec:method:setup}), computed with the Gaussian likelihood defined in Eq.~\ref{eq:Gaussian_likelihood}. Finally, we explore the source of the $f_{R0}\text{-}\beta$ degeneracy by examining the behaviour of theoretical predictions of the power spectrum across the parameter space (Section~\ref{sec:Pofk_constraints:degeneracy}).

\subsection{Shot noise estimation}\label{sec:Pofk_constraints:shot_noise}

In order to compute the likelihood of the mock galaxy power spectrum, $P_{\rm{g}}^{\rm{obs}}(k)$, given the parameters $\theta = (\beta,|f_{R0}|)$, we must first subtract the shot noise contribution from $P_{\rm{g}}^{\rm{obs}}(k)$.

We estimate the shot noise, $P_{\rm{g}}^{\rm{shot}}(k)$, as:
\begin{equation}\label{eq:shot_noise}
    P_{\rm{g}}^{\rm{shot}}(k) = \langle P_{\rm{g}}(k) \rangle_{N_{\rm{mock}}} -  P_{\rm{g}}^{\rm{fid}}(k). 
\end{equation}
Here, $P_{\rm{g}}^{\rm{fid}}(k)$ is the power spectrum of the noiseless fiducial galaxy number counts field, $n_{\rm{g}}^{\rm{fid}}$, and $\langle P_{\rm{g}}(k)\rangle_{N_{\rm{mock}}}$ represents the ensemble average of all $N_{\rm{mock}}=1024$ independent Poisson-sampled realisations. 

The left panel of Figure~\ref{fig:Pofk_shot_noise_demonstration} presents the power spectra for our GR fiducial model and demonstrate the impact of shot noise. The noiseless fiducial power spectrum, $P_{\rm{g}}^{\rm{fid}}(k)$ (black line) is compared against the power spectrum measured from the noisy mock data field, $P_{\rm{g}}^{\rm{obs}}(k)$ (dashed line), and the resulting shot-subtracted power spectrum, $P_{\rm{g}}^{\rm{ss}}(k) = P_{\rm{g}}^{\rm{obs}}(k) - P_{\rm{g}}^{\rm{shot}}(k)$ (dotted line). By subtracting shot noise from the mock observed power spectrum, we effectively recover the underlying theoretical prediction. 

The right panel of Figure~\ref{fig:Pofk_shot_noise_demonstration} shows the correlation matrix estimated from the ensemble of $N_{\rm{mock}}=1024$ shot-subtracted power spectra, $\{ P_{\rm{g}}^{\rm{ss},(j)}(k) \}_{j=1}^{N_{\rm{mock}}}$, where $P_{\rm g}^{\rm{ss},(j)}(k) \equiv P_{\rm{g}}^{(j)}(k) - P_{\rm{g}}^{\rm{shot}}(k)$ denotes the shot–subtracted power spectrum of the $j$-th noisy realisation. The resulting correlation matrix is strongly diagonal, as expected for largely independent Fourier modes in Gaussian fields, but exhibits non-negligible off-diagonal correlations at small $k$. These correlations are a direct consequence of the non-linear gravitational evolution of the cosmic web, which induces mode-coupling.

\subsection{Constraints from $P_g(k)$}\label{sec:Pofk_constraints:results}

Using the shot-subtracted power spectrum, $P_{\rm{g}}^{\rm{ss}}(k)$, we perform parameter inference on $(\beta,|f_{R0}|)$ following the pipeline detailed in Section~\ref{sec:method:param_estimation}. Figure \ref{fig:n_g_constraint} shows the resulting posterior distributions for our two fiducial cases. For the GR fiducial (left panel), the power spectrum analysis yields constraints on both $\beta$ and $|f_{R0}|$, however, a distinct degeneracy is evident: an increase in clustering power driven by a stronger fifth force (larger $|f_{R0}|$) can be compensated by a reduction in galaxy bias (smaller $\beta$). While the true GR value ($|f_{R0}|=0$) is consistent with the posterior distribution within our prior bounds, the 95\% confidence level allows for values up to $\log_{10}|f_{R0}| < -5.698$. For $\beta$, we obtain a 68\% confidence interval of $\beta = 0.771^{+0.005}_{-0.004}$, which is consistent with the true value of $\beta=0.770$. These constraints are obtained by imposing the prior boundaries of $\beta \in [0.75, 0.80]$ and $\log_{10}|f_{R0}| \in [-8,-4]$.
Since the GR limit is represented by negative infinity in logarithmic space, the posterior distribution tail is elongated toward increasingly negative values of $\log_{10}|f_{R0}|$.

For the F6 fiducial case (right panel), where the true value is $\log_{10}|f_{R0}|=-6$, the posterior distribution is centred close to the true values of $(\beta,|f_{R0}|)$ but remains very broad due to the same degeneracy observed in the GR fiducial case ($i.e.,$ the effects on galaxy clustering due to an increase in $|f_{R0}|$ can be compensated by a decrease in $\beta$, and vice versa). The analysis correctly prefers a non-zero $|f_{R0}|$, but it cannot distinguish the true value from a wide range of alternatives along the degeneracy direction. In fact, the $-2\sigma$ limit of $|f_{R0}|$ is pushed right against the lower boundary of our prior range showing that GR would remain allowed in such an inference. The recovered parameters within the 68\% confidence level are $\log_{10}|f_{R0}| = -6.223^{+1.032}_{-0.411}$ and $\beta = 0.766^{+0.007}_{-0.004}$, obtained by imposing the priors of $\beta \in [0.74, 0.79]$ and $\log_{10}|f_{R0}| \in [-8, -4]$. 

\subsection{Exploring the $\beta-|f_{R0}|$ degeneracy}\label{sec:Pofk_constraints:degeneracy}

We next seek to explore the origin of the degeneracy between $|f_{R0}|$ and $\beta$ that is inherent within the power spectrum of the galaxy number counts field. Figure~\ref{fig:Pofk_2sigma} presents the predicted galaxy power spectra for specific $(\beta,|f_{R0}|)$ combinations along the $1\sigma$ posterior boundaries in Figure~\ref{fig:n_g_constraint}, shown relative to the fiducial theoretical prediction, $P_{\rm{g}}^{\rm{fid}}(k)$ (GR fiducial: left panel; F6 fiducial: right panel). To characterise the observed degeneracy between $\beta$ and $|f_{R0}|$, we select combinations of $(\beta,|f_{R0}|)$ that isolate the competing effects of modified gravity and galaxy bias on clustering amplitudes, which are identified by either a star, triangle or circle (detailed below). The grey shaded regions in Figure~\ref{fig:Pofk_2sigma} represent the $1\sigma$ uncertainties on the fiducial power spectrum, estimated from the diagonal of the covariance matrix of $N_{\rm{mock}}=1024$ independent Poisson-sampled realisations. The models at the $1\sigma$ bounds have distinct power spectra, but the statistical uncertainty makes them difficult to distinguish.

Each of the predicted galaxy power spectra in Figure~\ref{fig:Pofk_2sigma} are computed by first generating the predicted galaxy number counts from the $\delta_{\rm{DM}}$ field of the chosen test gravity model, using the specified $\beta$, while setting the remaining galaxy bias parameters to those of the fiducial model ($N=0.19$, $\rho=1.61$ and $\epsilon=0.09$). Note that the tuning of $N(\beta, |f_{R0}|)$ does not take place, as the power spectrum does not depend on $N$ due to the cancellation in Eq.~\ref{eq:delta_g}. The predicted power spectrum is then computed from the number counts field using \textsc{Pylians}, as described in Section~\ref{sec:method:setup}. 

\paragraph*{Isolating galaxy bias effects.} 
Models marked with filled or hollow stars in both Figure~\ref{fig:n_g_constraint} and Figure~\ref{fig:Pofk_2sigma} represent models with increased or decreased galaxy bias, respectively. These are evaluated at the $\rm{\pm1\sigma}$ bounds of $\beta$ presented in Section~\ref{sec:Pofk_constraints:results}, with the fiducial gravity strength: $(\beta^{\rm{\pm1\sigma}},|f_{R0}|^{\rm{fid}}=0)$ for the GR fiducial case and $(\beta^{\rm{\pm1\sigma}},|f_{R0}|^{\rm{fid}}=10^{-6})$ for the F6 fiducial case. Note that, for the GR fiducial, the stars in Figure~\ref{fig:n_g_constraint} are positioned at our lower prior bound for $|f_{R0}|$ rather than the true value, as $\log_{10}|f_{R0}|=-\infty$ lies outside the sampled range. These $\beta^{\rm{\pm1\sigma}}$ values are taken directly from the marginalised statistics output of \textsc{emcee}. 

For both the GR case (left panel of Figure~\ref{fig:Pofk_2sigma}) and F6 fiducial case (right panel), increasing $\beta$ (filled stars) boosts power across all scales relative to the fiducial model due to the increase in galaxy clustering. This boost in power is scale-dependent given Eq.~\ref{eq:bias}. The opposite effect is true for decreasing beta (hollow stars).

\paragraph*{Isolating modified gravity effects.} 
Models marked with triangles in both Figure~\ref{fig:n_g_constraint} and Figure~\ref{fig:Pofk_2sigma} represent those with increased gravity strength, evaluated at the fiducial $\beta$ value: $(\beta^{\rm{fid}}=0.77,|f_{R0}|^{\rm{+1\sigma}})$ for both the GR and F6 fiducial case. These $|f_{R0}|^{\rm{\pm1\sigma}}$ values are obtained by selecting the closest $|f_{R0}|$ value sampled within our likelihood pipeline ($10^{-8} \leq |f_{R0}| \leq 10^{-4}$, $\Delta\log_{10}|f_{R0}|=0.25$) to the marginalised statistics output from \textsc{emcee}. For the GR fiducial case, the closest sampled value is $\log_{10}|f_{R0}|^{+1\sigma} = -5.75$, while in the F6 fiducial case, it is $\log_{10}|f_{R0}|^{+1\sigma} = -5.5$. In Figure~\ref{fig:Pofk_2sigma}, we only show the power spectrum predictions associated with the upper limit of $|f_{R0}|$, as we do not obtain a lower limit on $|f_{R0}|$ in the GR fiducial case, and the lower limit in the F6 fiducial case lies extremely close to our lower prior bound, making the upper limit more illustrative of clear deviations from the fiducial model prediction. 

We see that increasing $|f_{R0}|$ (filled triangles) results in a scale-dependent enhancement of power for increasing $k$, given the scale-dependent nature of the fifth force in $f(R)$ gravity which becomes more efficient at enhancing structure formation on smaller scales. A similar effect is observed for the F6 scenario.

\paragraph*{Degeneracy-line exploration.} 
Finally, we investigate a combination of $(\beta,|f_{R0}|)$ that falls along the degeneracy line of the posterior distribution. These degeneracy-line models are chosen to be near the $1\sigma$ boundaries and are highlighted by a circle in Figure~\ref{fig:n_g_constraint}, to specifically explore why combinations of increased $|f_{R0}|$ and decreased $\beta$ result in degenerate power spectra that remain indistinguishable within our statistical uncertainties. For the GR fiducial scenario, we choose the point $(\beta^{\rm{deg}}=0.766, \log_{10}|f_{R0}|^{\rm{deg}}=-6)$, while for the F6 fiducial, we choose $(\beta^{\rm{deg}}=0.753, \log_{10}|f_{R0}|^{\rm{deg}}=-5.5)$

Figure~\ref{fig:Pofk_2sigma} shows that the degeneracy-line model (higher $|f_{R0}|$ and lower $\beta$, filled circles) produces a power spectrum that decreases on larger scales but increases on smaller scales. These considered models are within the $1\sigma$ statistical uncertainties of the theoretical prediction of the fiducial model and are therefore indistinguishable from the fiducial.



\section{Full-field Analysis}\label{sec:full_field_constraints}

We now apply our full field-level likelihood framework to the same fiducial mock datasets as presented in Section~\ref{sec:Pofk_constraints} (adopting a single realisation of COLA initial conditions and a single realisation of Poisson noise). This section focuses on exploiting the full spatial information content of the $n_{\rm{g}}^{\rm{obs}}$ field to overcome the limitations of two-point statistics regarding the degeneracies between modified gravity and galaxy bias. 

We first summarise the joint constraints on $(\beta,|f_{R0}|)$ obtained with the full $n_{\rm{g}}$ field in Section~\ref{sec:full_field_constraints:results}. Next, we seek to explore the physical origin of this additional constraining power by examining the contributions of specific galaxy number counts to the total likelihood, identifying the most informative voxels in driving the constraints (Section~\ref{sec:full_field_constraints:n_g_contributions}). The spatial distribution of these voxels is then investigated in Section~\ref{sec:full_field_constraints:classifications} by classifying the matter fields into distinct cosmic environments. Finally, we isolate the role of Fourier phase information in breaking parameter degeneracies in Section~\ref{sec:full_field_constraints:phase_only}.

\begin{table*}
    \centering
    \caption{ \textbf{Top:} Total log-likelihood ratio contributions, $\sum \Delta \ln \mathcal{L}$, as a function of the observed galaxy count, $n_{\rm{g}}^{\rm{obs}}$ for the GR fiducial model. The table shows the total number of voxels, $N_{\rm{vox}}$, that have each galaxy count and the summed log-likelihood ratio for models at the $2\sigma$ parameter boundaries: (i) increased galaxy bias $(\beta^{+2\sigma}=0.779, |f_{R0}|^{\rm{fid}}=0)$; (ii) decreased galaxy bias $(\beta^{-2\sigma}=0.764, |f_{R0}|^{\rm{fid}}=0)$; (iii) increased gravity strength $(\beta^{\rm{fid}}=0.770, \log_{10}|f_{R0}|^{\rm{+2\sigma}}=-6.75)$. The fiducial $\beta$ value is $\beta^{\rm{fid}}=0.770$. 
     \textbf{Bottom:} Same as the top table, but for the F6 fiducial model showing models at the $2\sigma$ parameter boundaries: (i) increased galaxy bias $(\beta^{+2\sigma}=0.776, \log_{10}|f_{R0}|^{\rm{fid}}=-6)$; (ii) decreased galaxy bias $(\beta^{-2\sigma}=0.763, \log_{10}|f_{R0}|^{\rm{fid}}=-6)$; (iii) increased gravity strength $(\beta^{\rm{fid}}=0.770, \log_{10}|f_{R0}|^{+2\sigma}=-5.75)$; (iv) decreased gravity strength $(\beta^{\rm{fid}}=0.770, \log_{10}|f_{R0}|^{-2\sigma}=-6.25)$. 
     The teal and purple boxes represent positive and negative total log-likelihood ratio contributions, respectively, where more negative values corresponds to a stronger exclusion of the model. 
    }
    \renewcommand{\arraystretch}{1.3}  
    \setlength{\tabcolsep}{8pt}        
    \begin{tabular}{||ll||cccccccccc||}
        \hline
              &       & \multicolumn{10}{c||}{ \boldmath{$n_{\rm{g}}^{\rm{obs}}$: GR fiducial} }  \\
        \cline{3-12}
              &       & \textbf{0} & \textbf{1} & \textbf{2} & \textbf{3} & \textbf{4} & \textbf{5} & \textbf{6} & \textbf{7} & \textbf{8} & \textbf{9} \\
        \hline
        \hline
            \multicolumn{2}{||c||}{\boldmath{$N_{\rm{vox}}$}} & 2028543 & 64499 & 3543 & 440 & 88 & 24 & 10 & 2 & 2 & 1 \\
        \hline
        \hline
        \multirow{3}{*}{\boldmath{$\sum \Delta\ln\mathcal{L}$}} 
           & $\beta^{+2\sigma}$ & \pos{117.2} & \nega{-195.7} & \pos{46.3} & \pos{17.0} & \pos{5.5} & \pos{2.1} & \pos{1.1} & \pos{0.3} & \pos{0.3} & \pos{0.2} \\
           & $\beta^{-2\sigma}$ & \nega{-90.2} & \pos{145.8} & \nega{-37.0} & \nega{-13.5} & \nega{-4.4} & \nega{-1.7} & \nega{-0.9} & \nega{-0.2} & \nega{-0.2} & \nega{-0.2} \\
           & $f_{R0}^{+2\sigma}$ & \pos{8.3} & \nega{-14.9} & \pos{3.6} & \pos{1.0} & \pos{0.4} & \pos{0.2} & 0.0& 0.0 & 0.0 & 0.0 \\
        \hline
        \hline
    \end{tabular} \\[2mm]
     \renewcommand{\arraystretch}{1.3}  
    \setlength{\tabcolsep}{8pt}        
    \begin{tabular}{||ll||cccccccccc||}
        \hline
              &       & \multicolumn{10}{c||}{\boldmath{$n_{\rm{g}}^{\rm{obs}}$: F6 fiducial}}  \\
        \cline{3-12}
              &       & \textbf{0} & \textbf{1} & \textbf{2} & \textbf{3} & \textbf{4} & \textbf{5} & \textbf{6} & \textbf{7} & \textbf{8} &  \textbf{9} \\
        \hline
        \hline
          \multicolumn{2}{||c||}{\boldmath{$N_{\rm{vox}}$}} & 2028607 & 64304 & 3672 & 435 & 94 & 27 & 6 & 6 & 1 & 0\\
        \hline
        \hline
        \multirow{4}{*}{\boldmath{$\sum \Delta\ln\mathcal{L}$}} 
           & $\beta^{+2\sigma}$  & \pos{93.1}   & \nega{-154.8} & \pos{38.6} & \pos{13.4} & \pos{4.6} & \pos{1.8} & \pos{0.5} & \pos{0.7} & \pos{0.1} & 0.0  \\
          & $\beta^{-2\sigma}$  & \nega{-92.2} & \pos{149.1} & \nega{-39.5} & \nega{-13.6} & \nega{-4.7} & \nega{-1.9} & \nega{-0.5} & \nega{-0.7} & \nega{-0.1} & 0.0 \\
           & $f_{R0}^{+2\sigma}$ & \pos{57.7}   & \nega{-101.7} & \pos{28.6} & \pos{8.8} & \pos{3.0} & \pos{1.0} & \pos{0.3} & \pos{0.3} & 0.0 & 0.0 \\
          & $f_{R0}^{-2\sigma}$ & \nega{-35.3} & \pos{56.8} & \nega{-18.3} & \nega{-5.4} & \nega{-1.5} & \nega{-0.6} & 0.0 & \nega{-0.2} & 0.0 & 0.0 \\
        \hline
    \end{tabular}
    \label{tab:F6_fid_n_g_and_sum_loglike}
        \label{tab:GR_fid_n_g_and_sum_loglike}
\end{table*}

\subsection{Constraints from $n_{\rm{g}}$}\label{sec:full_field_constraints:results}

The joint $(\beta,|f_{R0}|)$ constraints obtained with the full galaxy number counts field are shown in Figure~\ref{fig:n_g_constraint}. For the GR fiducial (left panel), the full-field analysis yields a much tighter constraint on both $\log_{10}|f_{R0}|$ and $\beta$ compared to the power spectrum analysis. Crucially, we see that the field-level analysis has broken the degeneracy between them that persists when considering $P_{\rm{g}}(k)$ alone. The posterior of $\log_{10}|f_{R0}|$ is pushed towards the lower bound of our sampled range, correctly favouring GR, with a 95\% upper limit of $\log_{10}|f_{R0}| < -6.760$, compared to $\log_{10}|f_{R0}| < -5.698$ from the $P_{\rm{g}}(k)$ analysis. A comparison of the marginalised statistics from \textsc{emcee} from the field-level analysis (this section) and $P_{\rm{g}}(k)$ analysis (Section~\ref{sec:Pofk_constraints}) is presented in Table~\ref{tab:Pofk_vs_n_g_summary}. 
The constraint on $\beta$ has slightly smaller uncertainties, with $\beta=0.770\pm0.003$ from $n_{\rm{g}}$ versus $\beta=0.771^{+0.005}_{-0.004}$ from $P_{\rm{g}}(k)$.

For the F6 fiducial case (right panel), we see a similarly striking improvement in the constraints on both $\log_{10}|f_{R0}|$ and $\beta$. The full-field analysis yields a tight, unbiased constraint on $\log_{10}|f_{R0}|$ that is centred on the true value of $\log_{10}|f_{R0}|=-6$, with a $1\sigma$ confidence interval of $\log_{10}|f_{R0}|=-5.958^{+0.100}_{-0.107}$. By contrast, the $P_{\rm{g}}(k)$ analysis produces a broad, highly degenerate posterior with $1\sigma$ parameter recovery of $\log_{10}|f_{R0}| = -6.223^{+1.032}_{-0.411}$. The galaxy bias parameter, $\beta$, is also recovered with higher precision ($\beta=0.770\pm0.003$ using $n_{\rm{g}}$ compared to $\beta=0.766^{+0.007}_{-0.004}$ using $P_{\rm{g}}(k)$; Table~\ref{tab:Pofk_vs_n_g_summary}).

This enhancement in the constraining power arises from the non-Gaussian information contained within the full 3D field. While different models can produce power spectra that are indistinguishable beyond the estimated uncertainties (Figure~\ref{fig:Pofk_2sigma}), they predict non-Gaussian features in the field including phases. The field-level Poisson likelihood is sensitive to these differences which impact the cosmic web in unique ways, enabling it to distinguish between the effects of modified gravity and galaxy bias in ways that two-point statistics cannot.

\subsection{$n_{\rm{g}}$ Likelihood Contributions}\label{sec:full_field_constraints:n_g_contributions}

Here, we aim to isolate the specific galaxy number counts that drive the constraining power of our full-field likelihood framework. To do this, we analyse the log-likelihood contribution from each individual voxel for test models corresponding to the $2\sigma$ boundaries of the posterior distribution: we compute the voxel-by-voxel log-likelihood difference of these $\pm2\sigma$ models relative to the fiducial, identifying which voxels contribute most strongly to excluding that model. These models are chosen as they represent the boundary of statistical detectability from the true model and allow us to isolate the effects of varying gravity strength and galaxy bias across four distinct scenarios. 
The $2\sigma$ boundaries of both $\beta$ and $|f_{R0}|$ are determined by identifying the closest sampled values within our inference pipeline to the \textsc{emcee} marginalised constraints (since we sample a total of 17 $|f_{R0}|$ strengths, not including GR, and a total of 55 $\beta$ values; see Section~\ref{sec:method:param_estimation}). In particular, we find that the ability to distinguish between models is driven by specific galaxy number counts: voxels with $n_{\rm{g}}^{\rm{obs}}\neq1$ are highly sensitive to models with lower galaxy bias or gravity strength, while voxels with $n_{\rm{g}}^{\rm{obs}}=1$ are sensitive to increases in either galaxy bias or gravity strength.

Table~\ref{tab:GR_fid_n_g_and_sum_loglike} present the total number of voxels, $N_{\rm{vox}}$, and the summed log-likelihood ratio contributions, $\Sigma \Delta \ln \mathcal{L}$ as a function of the galaxy number count, $n_{\rm{g}}^{\rm{obs}}$. The log-likelihood ratio is computed as:
\begin{equation}
    \Delta \ln \mathcal{L} = \ln\mathcal{L}(n_{\rm{g}}^{\rm{obs}} \mid \theta^{\rm{test}}) - \ln\mathcal{L}(n_{\rm{g}}^{\rm{obs}} \mid \theta^{\rm{true}}),
\end{equation}
where $\ln\mathcal{L}(n_{\rm{g}}^{\rm{obs}} \mid \theta^{\rm{test}})$ is the log-likelihood of the model and $\ln\mathcal{L}(n_{\rm{g}}^{\rm{obs}} \mid \theta^{\rm{true}})$ is the log-likelihood of a model with the closest sampled parameters to the fiducial values, $\theta^{\rm{true}} \sim (\beta^{\rm{fid}}, |f_{R0}|^{\rm{fid}})$. More negative values of $\Delta \ln \mathcal{L}$ correspond to stronger exclusion of the test model. 

In both fiducial scenarios, we see that voxels with $n_{\rm{g}}^{\rm{obs}}\neq1$ are the most sensitive to models with low bias, while voxels with $n_{\rm{g}}^{\rm{obs}}=1$ are most sensitive to increased galaxy bias and increased strength of gravity. For a test model with lower bias ($\beta^{\rm{-2\sigma}}$), the highest log-likelihood ratio contribution comes from the voxels in which $n_{\rm{g}}^{\rm{obs}}=1$ (highlighted in teal in Table \ref{tab:GR_fid_n_g_and_sum_loglike}, whereas all voxels with number counts $n_{\rm{g}}^{\rm{obs}}\neq1$ yield a negative log-likelihood ratio contribution (highlighted in purple in Table \ref{tab:GR_fid_n_g_and_sum_loglike}).

\begin{figure}[h]
    \centering
    \includegraphics[width=\linewidth]{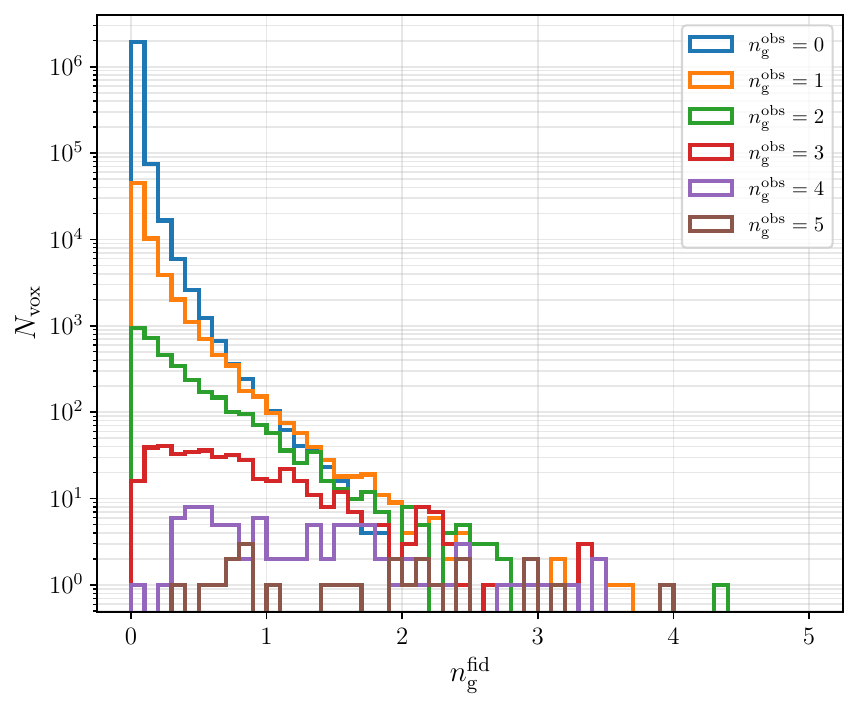}
    \caption{Distribution of the fiducial mean galaxy counts, $n_{\rm{g}}^{\rm{fid}}$, for given observed integer counts, $n_{\rm{g}}^{\rm{obs}}$, in the GR fiducial. The distributions show that voxels with $n_{\rm{g}}^{\rm{obs}}\geq1$ are predominantly drawn from low-density regions where $n_{\rm{g}}^{\rm{fid}}<1$. This confirms that the Poisson noise frequently increases the galaxy count within low-density regions of the fiducial model ($n_{\rm{g}}^{\rm{fid}}<1$) into the $n_{\rm{g}}^{\rm{obs}}\geq1$ regime, which accounts for the high sensitivity of our likelihood to these $n_{\rm{g}}^{\rm{obs}}=1$ and $n_{\rm{g}}^{\rm{obs}}\neq1$ voxels (see discussion in Section~\ref{sec:full_field_constraints:n_g_contributions}). }
    \label{fig:distribution_of_fid_n_g_GR}
\end{figure}

This occurs because a model with lower galaxy bias predicts fewer galaxies in higher density regions (larger $\delta_{\rm DM}$). Therefore, since we keep $N_{\rm tot}$ constant, the normalisation $N(\beta, |f_{R0}|)$ must increase to compensate the lower bias. This leads to an increased predicted galaxy number for fiducial number counts $n_{\rm{g}}^{\rm{fid}}<1$ and a decreased predicted galaxy number for $n_{\rm{g}}^{\rm{fid}}>1$. This is visualised in Figure~\ref{fig:distribution_of_fid_n_g_GR}, which shows the distribution of $n_{\rm{g}}^{\rm{fid}}$ for each observed count $n_{\rm{g}}^{\rm{obs}}$. Since there are more voxels with smaller $n_{\rm{g}}^{\rm{fid}}$, the voxels with $n_{\rm{g}}^{\rm{obs}} \geq 1$ preferentially originate from voxels with $n_{\rm{g}}^{\rm{fid}} < n_{\rm{g}}^{\rm{obs}}$ due to the Poisson noise, while the voxels with $n_{\rm{g}}^{\rm{obs}}=0$ can originate only from voxels with $n_{\rm{g}}^{\rm{fid}} \geq 0$. This provides a better fit (a positive log-likelihood ratio) in voxels where we observe $n_{\rm{g}}^{\rm{obs}}=1$, as lower bias increases $n_{\rm{g}}^{\rm{pred}} > n_g^{\rm{fid}}$ for these voxels, but a worse fit (a negative log-likelihood ratio) everywhere else (particularly in empty voxels where $n_{\rm{g}}^{\rm{obs}}=0$). We observe this same effect when considering the scenario of decreased gravity strength in the F6 fiducial. In this case, weaker gravity results in less pronounced dark matter overdensities and less extreme underdensities as compared to the fiducial model. This mimics the effect of decreased galaxy bias as both mechanisms suppress the clustering of the observed galaxy field relative to the underlying dark matter distribution.

Conversely, the opposite is true for a model with higher bias ($\beta^{\rm{+2\sigma}}$) or stronger gravity ($|f_{R0}|^{\rm{+2\sigma}}$). This makes observing voxels with $n_{\rm{g}}^{\rm{obs}}=1$ less probable, and results in large negative log-likelihood contributions from these voxels, which powerfully excludes these test models.

\subsection{Cosmic Web Classification}\label{sec:full_field_constraints:classifications}
To understand the origin of the additional constraining power in the field-level likelihood, we investigate which cosmic environments are most sensitive to the different models. In this section, we demonstrate that specific morphological environments contribute more to the total likelihood, therefore demonstrating their increased sensitivity to MG effects or variations in galaxy bias. This is an effect to which the power spectrum is insensitive.

We use the cosmic web classifier \textsc{PyCosmoMMF}\footnote{https://github.com/James11222/PyCosmoMMF} \citep{Sunseri_effects_2023} to classify our mock galaxy number count fields into four distinct morphological environments: voids, walls, filaments and clusters. This multi-scale morphological filter (MMF) algorithm, based on the \textsc{Nexus+} code \citep{Aragon-Calvo_multiscale_2007,Cautun_NEXUS_2013}, identifies structures based on the eigenvalues of the field tensor. More details of the \textsc{PyCosmoMMF} classification scheme can be found in \cite{Sunseri_effects_2023, Sunseri_power_2025}. We perform this classification on both the underlying DM density field, $\delta_{\rm{DM}}$, and the predicted $n_{\rm{g}}^{\rm{obs}}$ field.

We explore which cosmic environments host the voxels that are most effective at distinguishing between models. The critical question here is the spatial location of the highly informative $n_{\rm{g}}^{\rm{obs}}=1$ or $n_{\rm{g}}^{\rm{obs}}\neq1$ voxels
(see discussion in Section~\ref{sec:full_field_constraints:n_g_contributions}) that can exclude models with high bias/high modified gravity and low bias/weaker modified gravity, respectively. These are the voxels that serve to break the parameter degeneracy between galaxy bias and modified gravity. While two-point statistics are unable to isolate the spatial clustering of galaxies due to effects of an enhanced fifth force paired with a lower galaxy bias, or vice versa, the field-level likelihood exploits the fact that gravity and bias act on the spatial morphology of the cosmic web in different ways. We find that the most informative voxels 
reside in specific cosmic environments. Table~\ref{tab:voxel_count_in_classifications} shows, for both the GR fiducial (left) and F6 fiducial (right), the total number of voxels that have $n_{\rm{g}}=1$ or $n_{\rm{g}}\neq1$ within each cosmic classification of the $n_{\rm{g}}$ field (top row) and of the $\delta_{\rm{DM}}$ field (bottom row). We see that, for both the GR and F6 fiducial, the majority of voxels containing $n_{\rm{g}}\neq1$ reside in the voids with $n_{\rm{g}}$ very close to 0 as voids make up the majority of the Universe,
while the majority of the voxels containing $n_{\rm{g}}=1$ reside within the filaments.

\begin{figure*}
    \centering
    \includegraphics[width=\textwidth]{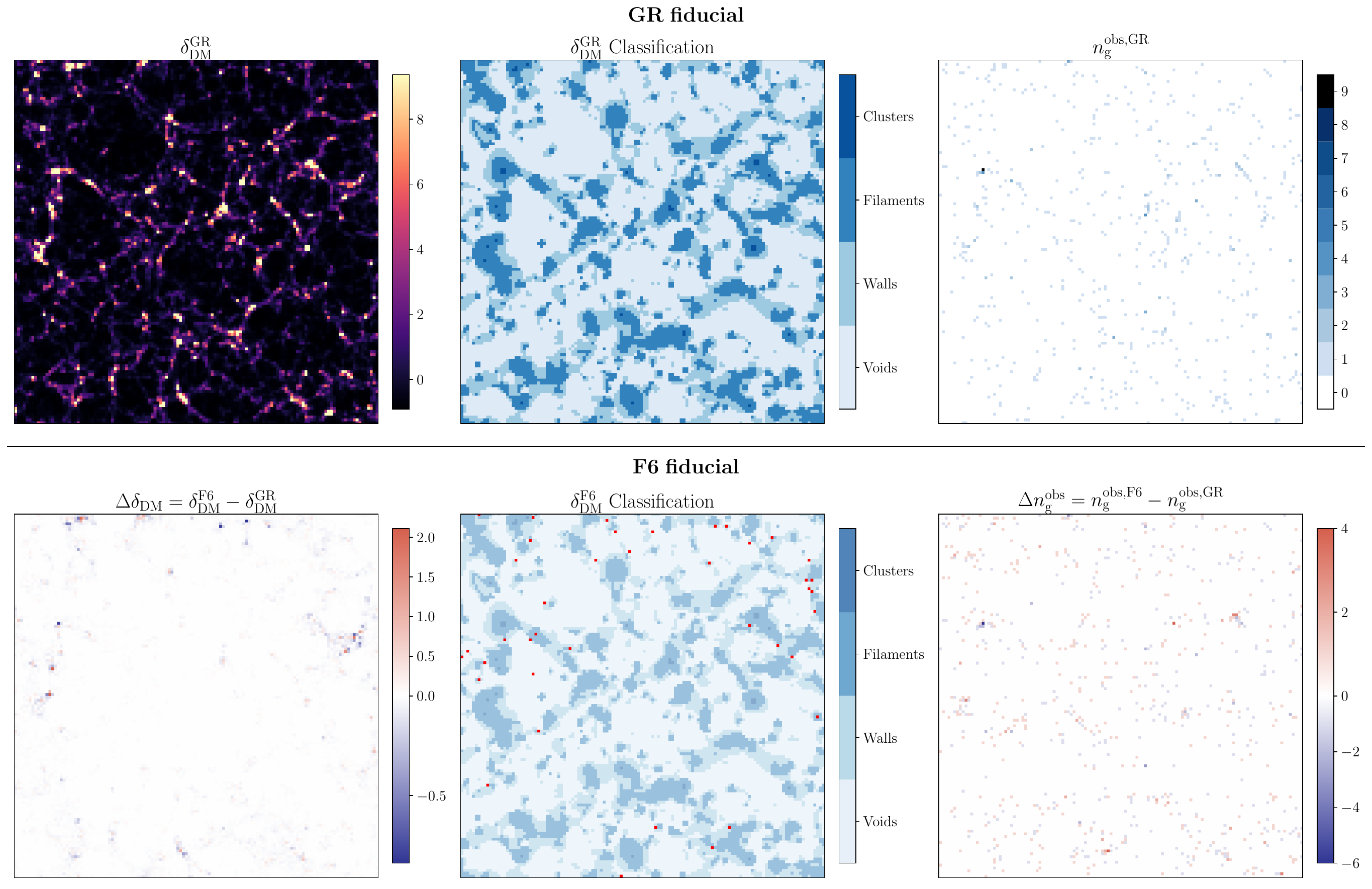}  
    \caption{Visualisation of a 2D slice through the simulation volume for GR and F6 fiducial models. \textbf{Top row:} The GR fiducial dataset, showing the DM overdensity, $\delta_{\rm{DM}}^{\rm{GR}}$ (left), the cosmic web classification according to $\delta_{\rm{DM}}^{\rm{GR}}$ (middle), and the corresponding galaxy number counts field $n_{\rm{g}}^{\rm{obs,GR}}$ (right). \textbf{Bottom row:} The F6 fiducial dataset, showing the difference in DM overdensity relative to GR, $\Delta \delta_{\rm{DM}}=\delta_{\rm{DM}}^{\rm{F6}} - \delta_{\rm{DM}}^{\rm{GR}}$ (left), the cosmic web classification according to $\delta_{\rm{DM}}^{\rm{F6}}$, with voxels differing from the GR classification being highlighted in red (middle), and the difference in galaxy counts, $\Delta n_{\rm{g}}^{\rm{obs}} = n_{\rm{g}}^{\rm{obs,F6}}-n_{\rm{g}}^{\rm{obs,GR}}$. The slice of our simulated volume has side length of 400 $h^{-1}$ Mpc and thickness of 3.125 $h^{-1}$ Mpc. This slice is chosen according to the location of the highest-density voxel within $n_{\rm{g}}^{\rm{obs,GR}}$ which has a maximum of $n_{\rm{g,max}}^{\rm{obs,GR}}=9$. The maximum galaxy count in the corresponding slice of the F6 dataset is $n_{\rm{g,max}}^{\rm{obs,F6}}=4$. Cosmic web classifications computed with \textsc{PyCosmoMMF} \citep{Sunseri_effects_2023}.  }
    \label{fig:slice_distributions}
\end{figure*}

Figure~\ref{fig:slice_distributions} shows a 2D slice through our simulations in GR (top row) and F6 (bottom row). The left panels show $\delta_{\rm{DM}}$ for the GR fiducial (top), and F6 relative to GR ($\Delta\delta_{\rm{DM}} = \delta_{\rm{DM}}^{\rm{F6}} - \delta_{\rm{DM}}^{\rm{GR}}$, bottom). The middle panels show the corresponding DM environmental classifications obtained with \textsc{PyCosmoMMF}; as the morphological differences between $\delta_{\rm{DM}}^{\rm{GR}}$ and $\delta_{\rm{DM}}^{\rm{F6}}$ are subtle in this particular slice, the F6 classifications plot is displayed with reduced opacity and voxels in which the cosmic environment classification is different to that of the GR fiducial are indicated in red. Finally, the right panels show the galaxy number counts field $n_{\rm{g}}^{\rm{obs}}$ in GR (top), and F6 relative to GR ($\Delta n_{\rm{g}} = n_{\rm{g}}^{\rm{F6}} - n_{\rm{g}}^{\rm{GR}}$, bottom). The slice is chosen according to the location of the highest-density $n_{\rm{g}}^{\rm{obs}}$ voxel in GR, which has a maximum value of $n_{\rm{g,max}}^{\rm{obs,GR}}=9$. The corresponding slice in the F6 model has a maximum count of $n_{\rm{g,max}}^{\rm{obs,F6}}=4$. The slice has a thickness of 3.125 $h^{-1}$ Mpc. 

\begin{table*}
\centering
\caption{Total number of voxels in which $n_{\rm{g}}^{\rm{obs}}=1$ and $n_{\rm{g}}^{\rm{obs}}\neq1$ (see Section~\ref{sec:full_field_constraints:n_g_contributions}) for each cosmic environment, as classified by the galaxy number counts field, $n_{\rm{g}}^{\rm{obs}}$ (top), and the underlying DM field, $\delta_{\rm{DM}}$ (bottom). \textbf{Left:} GR fiducial. \textbf{Right:} F6 fiducial. For both fiducial scenarios, the majority of voxels containing $n_{\rm{g}}^{\rm{obs}}\neq1$ reside within the voids, while the majority of those containing $n_{\rm{g}}^{\rm{obs}}=1$ reside within the filaments.}
\renewcommand{\arraystretch}{1.3}  
\setlength{\tabcolsep}{8pt}       
\begin{tabular}{||c||c|c|c|c||c|c|c|c||}
\hline
\multicolumn{1}{||c||}{} & \multicolumn{4}{c||}{\textbf{GR fiducial}} & \multicolumn{4}{c||}{\textbf{F6 fiducial}} \\
\hline
\multicolumn{1}{||c||}{\textbf{$n_{\rm{g}}$ classification}} & Voids & Walls & Filaments & Clusters & Voids & Walls & Filaments & Clusters \\
\hline
\hline
$n_{\rm{g}}^{\rm{obs}}\neq1$ & 1,032,818 & 434,147 & 565,503 & 185   & 1,029,684 & 440,025 & 562,759 & 380 \\
$n_{\rm{g}}^{\rm{obs}}=1$    & 28        & 19      & 64,449  & 3     & 23        & 8       & 64,248  & 25 \\
\hline
\multicolumn{9}{c}{} \\[-2mm] 
\hline
\multicolumn{1}{||c||}{\textbf{DM classification}} & Voids & Walls & Filaments & Clusters & Voids & Walls & Filaments & Clusters \\
\hline
\hline
$n_{\rm{g}}^{\rm{obs}}\neq1$ & 1,271,515 & 438,617 & 320,427 & 2,094  & 1,270,741 & 437,899 & 321,985 & 2,223 \\
$n_{\rm{g}}^{\rm{obs}}=1$    & 20,882    & 15,677  & 26,972  & 968    & 20,840    & 15,528  & 26,960  & 976 \\
\hline
\end{tabular}
\label{tab:voxel_count_in_classifications}
\end{table*}

\subsubsection{Galaxy field ($n_{\rm{g}}^{\rm{obs}}$) classification}
We first test whether classifying the observed galaxy count field, $n_{\rm{g}}^{\rm{obs}}$, into distinct cosmic environments yields information about the underlying physics. As shown in Table~\ref{tab:voxel_count_in_classifications}, the majority of voxels with $n_{\rm{g}}^{\rm{obs}}=1$ are located within filaments, while voxels with $n_{\rm{g}}^{\rm{obs}}\neq1$ are primarily located within voids. Therefore, classifying the galaxy field does not give us additional insight into the breaking of parameter degeneracies identified in Section~\ref{sec:full_field_constraints:n_g_contributions}. We therefore focus our analysis on the underlying DM density field.
However, cosmic-web classification based directly on the galaxy field can be performed on observational data, whereas the underlying dark-matter field must generally be inferred through reconstruction or constrained-simulation techniques; recent work using constrained simulations based on BORG has demonstrated how such reconstructions can be used to associate observed galaxies with their dark-matter density and cosmic-web environments while propagating reconstruction uncertainties~\citep{Catherine}.

\begin{figure*}
    \centering
    \begin{minipage}{\linewidth}
        \centering
        \captionof{table}{GR fiducial. For each $2\sigma$ boundary test model isolating the effects of either (i) increased galaxy bias, (ii) decreased galaxy bias, (iii) increased gravity strength, we identify which cosmic environments drive the exclusion of that model relative to the fiducial. Each entry reports the most negative summed log-likelihood ratio, $\sum \Delta \ln \mathcal{L}$, evaluated over all voxels belonging to the specified category. More negative values therefore correspond to stronger exclusion of the test model. The first data column lists the total contribution from the galaxy count mask alone ($n_{\rm{g}}^{\rm{obs}}=1$ or $n_{\rm{g}}^{\rm{obs}}\neq1$), reproducing the dominant contributions identified in Table~\ref{tab:GR_fid_n_g_and_sum_loglike} (purple cells). The following columns show the dominant contributions when grouping voxels by (i) $n_{\rm{g}}^{\rm{obs}}$ classification, (ii) $\delta_{\rm{DM}}$ classification, (iii) the combination of $n_{\rm{g}}^{\rm{obs}}$ mask and $\delta_{\rm{DM}}$ classification. The bold, coloured font indicate the environments that provide the strongest exclusion, namely $n_{\rm{g}}^{\rm{obs}}=1$ within voids for high bias and $n_{\rm{g}}^{\rm{obs}}=1$ within walls for high gravity. These are visualised in Figure~\ref{fig:GR_voids_walls_with_n_g==1}. }
        \renewcommand{\arraystretch}{1.3}  
        \begin{tabular}{||c|c||c|c|c|c|}
        \multicolumn{5}{c}{\textbf{GR fiducial}}\\
        \hline
         & $n_{\rm{g}}^{\rm{obs}}$ mask: $\sum \Delta \ln \mathcal{L}$ 
         & $n_{\rm{g}}^{\rm{obs}}$ classification: $\sum \Delta \ln \mathcal{L}$  
         & $\delta_{\rm{DM}}$ classification: $\sum \Delta \ln \mathcal{L}$
         & $\delta_{\rm{DM}}$ classification + $n_{\rm{g}}^{\rm{obs}}$ mask: $\sum \Delta \ln \mathcal{L}$  \\ 
         
        \hline 
        \hline
         (i) $\beta^{+2\sigma}$ 
         & $n_{\rm{g}}^{\rm{obs}}=1$: -195.7 
         & Filaments: -216.0 
         & \begin{tabular}{c}
            Walls: -2.0 \\
            Filaments: -2.1
           \end{tabular}
         & \orange{\textbf{Voids} + $\boldsymbol{n_{\rm{g}}^{\rm{obs}}=1}$}: -227.6 \\
        \hline
         (ii) $\beta^{-2\sigma}$ 
         & $n_{\rm{g}}^{\rm{obs}} \neq 1$: -148.3 
         & Voids: -122.2 
         & \begin{tabular}{c}
            Voids: -1.4 \\
            Filaments: -1.0 \\
            Clusters: -0.9
           \end{tabular}
         & Voids + $n_{\rm{g}}^{\rm{obs}} \neq 1$: -176.5 \\
        \hline
        (iii) $|f_{R0}|^{+2\sigma}$ 
        & $n_{\rm{g}}^{\rm{obs}}=1$, -14.9 
        & Filaments: -6.3  
        & \begin{tabular}{c}
            Walls: -0.8 \\
            Filaments: -0.4 \\
            Voids: -0.3
           \end{tabular}
        & \begin{tabular}{c}
            Voids + $n_{\rm{g}}^{\rm{obs}}=1$: -6.4 \\
            \orange{\textbf{Walls} + $\boldsymbol{n_{\rm{g}}^{\rm{obs}}=1}$}: -6.9
           \end{tabular} \\
        \hline
        \end{tabular}
        \label{tab:GR_fid_classification_summary}
    \end{minipage}
    \vspace{2em} 
    \begin{minipage}{\linewidth}
        \centering 
        \includegraphics[width=1\linewidth]{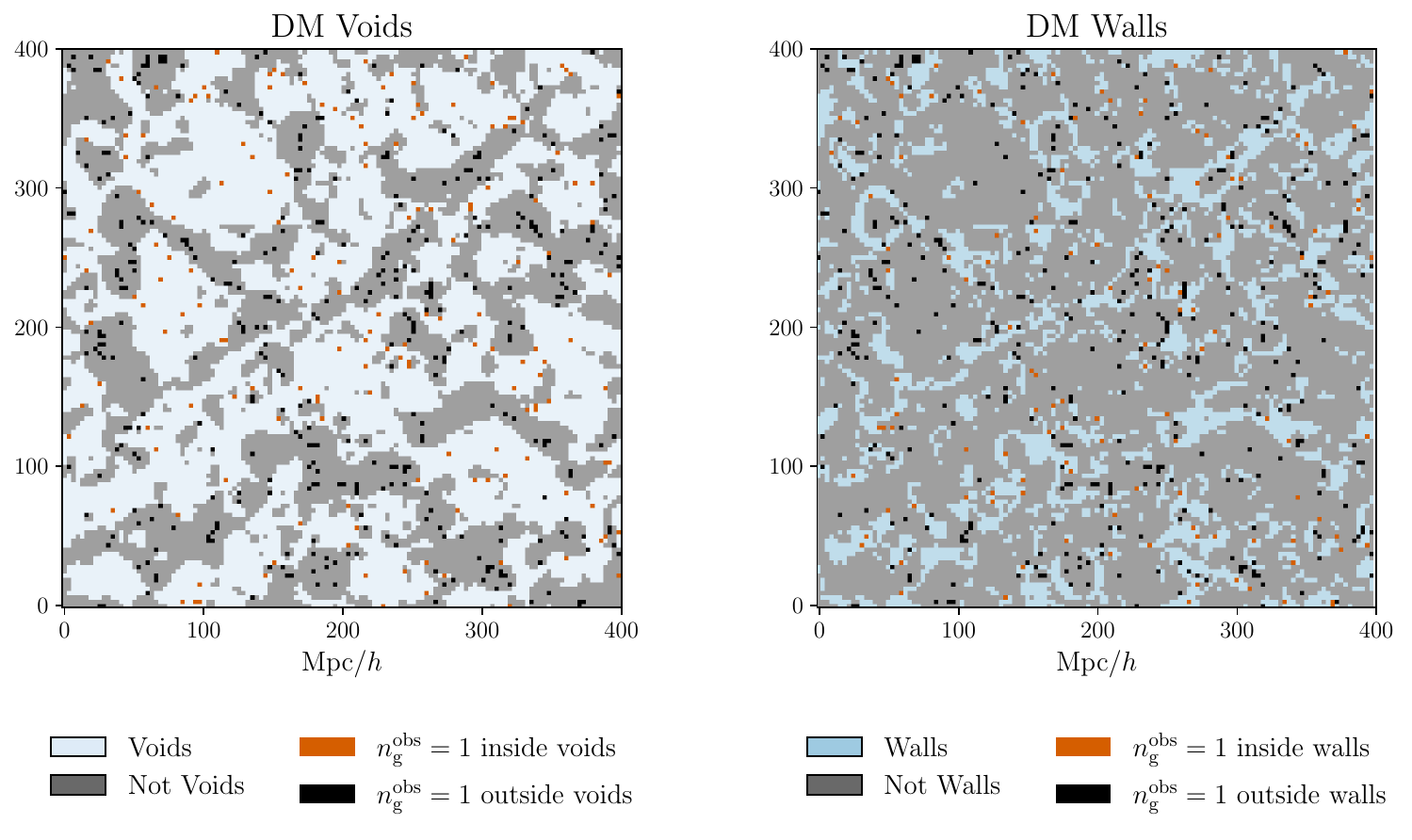}
        \caption{GR fiducial. Slice of the DM density field showing the classification of voxels as voids (left) and walls (right). Lighter regions show the respective cosmic web classification, while darker shaded regions are outside of these structures. Galaxies with $n_{\rm{g}}^{\rm{obs}}=1$ inside the structures are shown in orange, and voxels with $n_{\rm{g}}=1$ outside voids are shown in black. This figure demonstrates the voxels contributing to the likelihood for the cells highlighted in Table~\ref{tab:GR_fid_classification_summary}, $i.e.$, $n_{\rm{g}}^{\rm{obs}}=1$ within voids (left), $n_{\rm{g}}^{\rm{obs}}=1$ within walls (right). Cosmic web classifications computed with \textsc{PyCosmoMMF} \citep{Sunseri_effects_2023}.}
        \label{fig:GR_voids_walls_with_n_g==1}
    \end{minipage}
\end{figure*}


\begin{figure*}
    \centering

    \begin{minipage}{\linewidth}
        \centering
        \captionof{table}{Same analysis as in Table~\ref{tab:GR_fid_classification_summary} but for the F6 fiducial, including the additional test case of (iv) decreased gravity strength ($|f_{R0}|^{-2\sigma}$). In this case, the first data column lists the dominant contributions identified in Table~\ref{tab:F6_fid_n_g_and_sum_loglike} (purple cells).}
    
        \renewcommand{\arraystretch}{1.3}
        \begin{tabular}{||c|c||c|c|c|c||}
        \multicolumn{5}{c}{\textbf{F6 fiducial}}\\
        \hline
         & $n_{\rm{g}}^{\rm{obs}}$ mask: $\sum \Delta \ln \mathcal{L}$ 
         & $n_{\rm{g}}^{\rm{obs}}$ classification: $\sum \Delta \ln \mathcal{L}$ 
         & $\delta_{\rm{DM}}$ classification: $\sum \Delta \ln \mathcal{L}$ 
         & $\delta_{\rm{DM}}$ classification + $n_{\rm{g}}^{\rm{obs}}$ mask: $\sum \Delta \ln \mathcal{L}$  \\
        \hline
        \hline
        
        $\beta^{+2\sigma}$ 
        & $n_{\rm{g}}^{\rm{obs}}=1$: -154.8 
        & Filaments: -170.4 
        & \begin{tabular}{c}
            Walls: -1.0 \\
            Filaments: -1.7
          \end{tabular}
        & $n_{\rm{g}}^{\rm{obs}}=1$ + Voids: -178.0 \\
        \hline  
        
        $\beta^{-2\sigma}$ 
        & $n_{\rm{g}}^{\rm{obs}} \neq 1$: -153.2 
        & Voids: -123.2  
        & \begin{tabular}{c}
            Voids: -1.1 \\
            Clusters: -2.5
          \end{tabular}
        & Voids: -177.4 \\
        \hline
        
        $f_{R0}^{+2\sigma}$ 
        & $n_{\rm{g}}^{\rm{obs}}=1$: -101.7 
        & Filaments: -76.5 
        & \begin{tabular}{c}
            Walls: -1.6 \\
            Filaments: -2.8
          \end{tabular}
        & \begin{tabular}{c}
            Voids: -75.5 \\
            $n_{\rm{g}}^{\rm{obs}}=1$ + Walls: -48.4
          \end{tabular} \\
        \hline
        
        $f_{R0}^{-2\sigma}$ 
        & $n_{\rm{g}}^{\rm{obs}}\neq1$: -61.3 
        & \begin{tabular}{c}
            Voids: -19.4 \\
            Walls: -15.5
          \end{tabular}
        & Filaments: -3.1  
        & \begin{tabular}{c}
            Voids: -41.2 \\
            Walls: -28.4
          \end{tabular} \\
        \hline
        
        \end{tabular}
        \label{tab:F6_fid_classification_summary}
    \end{minipage}
\end{figure*}

\subsubsection{Dark matter ($\delta_{\rm{DM}}$) classification}
In contrast, by classifying the underlying DM density field into distinct cosmic environments and pairing this with the $n_{\rm{g}}^{\rm{obs}}=1$ and $n_{\rm{g}}^{\rm{obs}}\neq1$ criteria, we can identify the specific cosmic environments driving the constraints. \footnote{This focus on the underlying dark-matter field is also supported by~\citet{Catherine}, who find that dark-matter density is more strongly correlated with observed galaxy properties than distance to the nearest filament is, suggesting that density retains more of the environmentally relevant information than cosmic-web location alone.} We present these findings in Table~\ref{tab:GR_fid_classification_summary} and Figure~\ref{fig:GR_voids_walls_with_n_g==1} for the GR fiducial case and in Table~\ref{tab:F6_fid_classification_summary} for the F6 fiducial case. Note that we do not show the combination of $n_{\rm{g}}^{\rm{obs}}$ mask and $n_{\rm{g}}^{\rm{obs}}$ classification as the majority of these $n_{\rm{g}}^{\rm{obs}}=1$ and $n_{\rm{g}}^{\rm{obs}}\neq1$ voxels live within the filaments and voids, respectively, so we do not gain any additional information. The exact $2\sigma$ values are taken to be the closest sampled values within our inference pipeline to the marginalised \textsc{emcee} constraints.  We observe that the voxels most capable of discriminating between galaxy bias and modified gravity effects are located primarily within the voids and walls. Specifically, When testing a high-bias or high-$|f_{R0}|$ model, the dominant constraining power (the most negative log-likelihood) comes from voxels with $n_{\rm{g}}^{\rm{obs}}=1$ that reside in underdense regions, specifically DM voids and walls. An enhanced bias or a fifth force makes $n_{\rm{g}}^{\rm{obs}}=1$ voxels in these underdense regions even more unlikely than in the fiducial model. 

While the DM classification alone does not give useful information, it is the pairing with the field-level $n_{\rm{g}}^{\rm{obs}}$ masks that can isolate the physical effects. In the DM voids, voxels containing $n_{\rm{g}}^{\rm{obs}}=1$ are highly sensitive to the effects of galaxy bias, while the $n_{\rm{g}}^{\rm{obs}}=1$ voxels within the walls are isolating the effects of MG. This shows that even though the effects of galaxy bias and modified gravity are similar when considering the $n_g$ likelihood contribution (Tables~\ref{tab:GR_fid_n_g_and_sum_loglike}) the difference of these two effects appear when we consider the spatial morphology of the cosmic web. As previously demonstrated, for example in \cite{Pisani_cosmic_2019}, voids and their surrounding walls serve as a powerful probe for testing departures from GR. We show where these $n_{\rm{g}}^{\rm{obs}}$ voxels are located within the voids and walls in Figures~\ref{fig:GR_voids_walls_with_n_g==1} for the GR fiducial models. 


\begin{figure*}
    \centering
    \includegraphics[width=\textwidth]{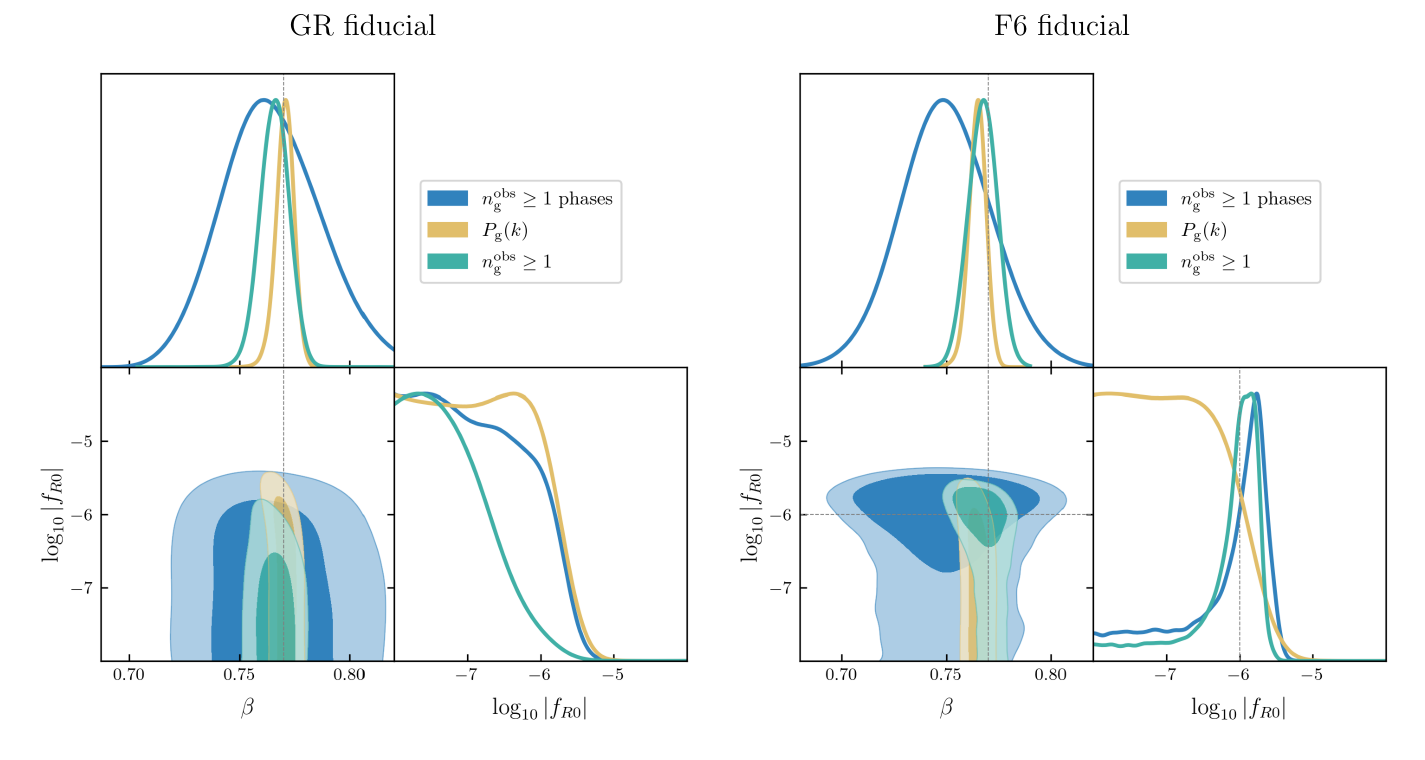}
    \caption{Constraints on $\beta$ and $|f_{R0}|$ obtained by isolating the information content of the Fourier phases for the GR fiducial model (left) and the F6 fiducial model (right). We compare these phase-only constraints (blue) to amplitude-only constraints from the power spectrum using the full field, $P_{\rm{g}}(k)$ (yellow), and combined phase + amplitudes constraints obtained with the full-field analysis of $n_{\rm{g}}^{\rm{obs}}\geq1$ voxels (green). As discussed in Section~\ref{sec:full_field_constraints:phase_only}, we impose a threshold of $n_{\rm{g}}^{\rm{obs}}\geq1$ due to numerical errors that occur when replacing the Fourier amplitudes. The phase information is critical for breaking the degeneracies between $|f_{R0}|$ and $\beta$ that are inherent in the $P_{\rm{g}}(k)$ analysis. GR fiducial: phase-only prior range of $\beta \in [0.70,0.83]$. F6 fiducial: phase-only prior range of $\beta \in [0.68,0.82]$. Prior ranges of $P_{\rm{g}}(k)$ and $n_{\rm{g}}\geq1$ are the same as those in Figure~\ref{fig:n_g_constraint}. Phase-only and full-field analyses use the truncated Poisson likelihood in Eq.~\ref{eq:truncPoisson_likelihood}, while the $P_{\rm{g}}(k)$ analysis uses the Gaussian likelihood in Eq.~\ref{eq:Gaussian_likelihood}. 
    }
    \label{fig:phase_only}
\end{figure*}

\subsection{Role of Fourier Phases}\label{sec:full_field_constraints:phase_only}
A key advantage of the full-field analysis is that it naturally incorporates Fourier phase information, which is absent from two-point statistics. To explicitly demonstrate the importance of Fourier phase information on model exclusion, we conduct a "phase-only" test. In this test, we construct a new set of theoretically predicted fields, $n_{\rm{g}}^{\rm{pred}}$, sampled over a 2D grid of $(\beta, |f_{R0}|)$ as described in Section~\ref{sec:method:param_estimation}. To isolate the contribution of spatial morphology, we impose that the mock observed field and all predicted fields share identical Fourier amplitudes, but retain their original, model-specific phases, which encode the spatial morphology of the galaxy number counts field. 

We achieve this test by first imposing a threshold on the mock dataset of $n_{\rm{g}}^{\rm{obs}} \geq 1$ and applying this mask to the predicted fields $n_{\rm{g}}^{\rm{pred}}$. We impose a threshold of $n_{\rm{g}}^{\rm{obs}}\geq1$ due to numerical errors that arise when replacing the Fourier amplitudes of voxels that contain no galaxies. Then, by taking the Fourier transform of both $n_{\rm{g}}^{\rm{obs}}$ and $n_{\rm{g}}^{\rm{pred}}$, we replace the Fourier amplitudes of all fields with those of the noiseless fiducial field $n_{\rm{g}}^{\rm{fid}}$, to ensure that the original Fourier amplitude information of the fields cannot be used to constrain parameters. This choice of replacing the Fourier amplitudes of the observed and predicted fields with those of the fiducial field is purely for consistency, as we do not add noise to the theoretical predictions $n_{\rm{g}}^{\rm{pred}}$. Finally, by performing the inverse Fourier transform back into real space, we obtain a set of $n_{\rm{g}}$ fields that preserve their original spatial morphology via their original Fourier phases, while removing all amplitude information that would be considered in the power-spectrum-only analysis, by imposing that they all share the same Fourier amplitudes. Because we have imposed a threshold on $n_{\rm{g}}^{\rm{obs}}$ to perform this phase-only test, we conduct our likelihood inference using the truncated Poisson likelihood in Eq.~\ref{eq:truncPoisson_likelihood}.


We then repeat the likelihood analysis using these phase-only fields. We show the results of this phase-only test in Figure~\ref{fig:phase_only}, where we compare the phase-only constraint (blue contours), the amplitudes-only constraint (the $P_{\rm{g}}(k)$ analysis presented in Figure~\ref{fig:n_g_constraint}, which uses the full field but with a different Poisson seed, yellow contours), and the constraint from both Fourier amplitudes and phases (the full $n_{\rm{g}}$ field) in voxels where $n_{\rm{g}}^{\rm{obs}}\geq1$ (green contours). The left panel shows the results of the GR fiducial model, and the right panel shows the results of the F6 fiducial model. When considering only the phases, we lose a lot of the constraining power on both $\beta$ and $|f_{R0}|$ than when considering the full $n_{\rm{g}}$ field, but we see that the phases alone are enough to obtain a more accurate parameter recovery of $|f_{R0}|$. In the F6 case in particular, although the $P_{\rm{g}}(k)$ analysis using the full field without the threshold provides a much tighter constraint on $\beta$, we are unable to distinguish between GR and $|f_{R0}| > 0$ when considering the $P_{\rm{g}}(k)$ analysis. The green contour (constraint using both phase and amplitude information contained within the $n_{\rm{g}}^{\rm{obs}}\geq1$ field) highlights the power of using a combination of phase and amplitude information. While the phases alone do not provide a tight constraint, they provide information that is complementary to the amplitudes. 

As is more evident in the F6 fiducial case, they are able to break the degeneracy between $|f_{R0}|$ and $\beta$ that is evident in the $P_{\rm{g}}(k)$ analysis, and can correctly identify the true value of $\log_{10}|f_{R0}|$. When this phase information is combined with amplitude information in the full-field analysis of $n_{\rm{g}}^{\rm{obs}}\geq1$ voxels, we obtain a tighter constraint on both parameters. This explicitly shows that the superior performance of the field-level approach stems from its ability to harness simultaneously both amplitude and phase information, which is lost in $P_{\rm{g}}(k)$. 

Theoretically, it is interesting that modified gravity affects phase correlations, and that this information alone can provide constraints. However, this analysis is done with a fixed initial seed which is assumed to be known. When applied to real data, marginalising over the initial phases is expected to eliminate the constraining power from phase correlations. Any remaining constraints would then arise from the non-Gaussianity of the Fourier amplitudes probed by the $n_{\rm g}$ analysis, which is indeed the dominant source of the constraining power, as evidenced by the substantial difference between the phase-only and full-field constraints shown in Figure~\ref{fig:phase_only}.


\section{Tests of Robustness}\label{sec:robustness}

To validate the reliability and stability of our field-level inference framework, we perform a series of tests to assess the robustness and stability of our framework under varied different conditions. Specifically, we study the sensitivity of our posterior constraints to the choice of thresholding applied to the mock galaxy number counts field, $n_{\rm{g}}^{\rm{obs}}$ (Section~\ref{sec:robustness:thresholding}), and furthermore, assess the impact of variations in the realisation of Poisson noise and in the initial conditions of the COLA simulations generating the DM overdensity field (Section~\ref{sec:robustness:Poisson_seeds}). These tests are designed to demonstrate that our results are not artefacts of a specific data realisation, but are instead representative of the superior constraining power of our field-level inference method.

\begin{figure*}
    \centering
    \includegraphics[width=\textwidth]{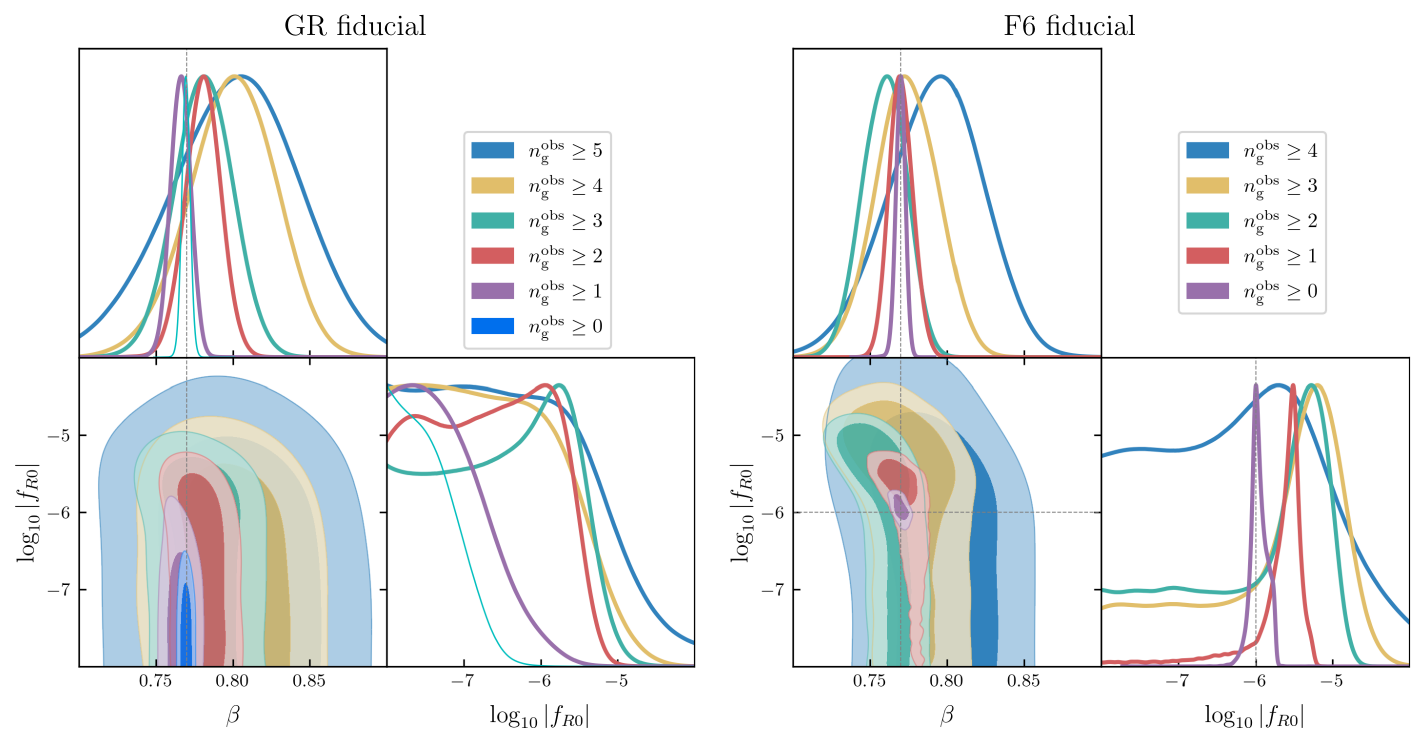}
    \caption{Constraint on $\beta$ and $|f_{R0}|$ derived from the full $n_{\rm{g}}^{\rm{obs}}$ field and cases where a threshold, $T$, is applied. Left: GR fiducial, where thresholds on $n_{\rm{g}}^{\rm{obs}}$ are $T=0,1,2,3,4,5$. Right: F6 fiducial, where thresholds on $n_{\rm{g}}^{\rm{obs}}$ are $T=0,1,2,3,4$. The full-field constrains utilise the Poisson likelihood (Eq.~\ref{eq:Poisson_likelihood}) with a prior range of $\beta\in[0.75,0.80]$ ($\beta\in[0.74,0.79]$) for GR (F6). Thresholded constraints use the truncated Poisson likelihood (Eq.~\ref{eq:truncPoisson_likelihood}), with prior ranges on $\beta$ of $\beta \in [0.70,0.90]$ for both GR and F6 fiducial.}
    \label{fig:thresholding_test}
\end{figure*}

\subsection{Thresholding $n_{\rm{g}}^{\rm{obs}}$}\label{sec:robustness:thresholding}

We first investigate how the constraints depend on the number of galaxies included within the analysis. We repeat our likelihood analysis including only voxels where the mock galaxy counts data field exceeds a certain threshold, $T$, $n_{\rm{g}}^{\rm{obs}} \geq T$, for various values of $T$. Within our pipeline, a mask is created by applying the threshold to the $n_{\rm{g}}^{\rm{obs}}$, and this mask is consistently applied to all predicted fields, $n_{\rm{g}}^{\rm{pred}}(\beta, |f_{R0}|)$ ($i.e.$, the mask is based on the mock data). We have tested and confirmed that in the high threshold limit, the Poisson likelihood results agree with field-level analysis using a Gaussian likelihood, as expected. The prior ranges used for these thresholded cases are $\beta\in[0.70, 0.90]$ for both fiducial scenarios.

Figure~\ref{fig:thresholding_test} shows the resulting parameter constraints for the GR (left) and F6 (right) fiducial models. As expected, increasing the threshold $T$ progressively discards information, leading to a systematic broadening of the posterior contours and a weaked constraining power on both $|f_{R0}|$ and $\beta$. However, even with a threshold of $n_{\rm{g}}^{\rm{obs}}>5$, our field-level approach manages to successfully recover an upper limit on $\log_{10}|f_{R0}|$. This demonstrates that while the information gain from low-density regions are important (as discussed in Section~\ref{sec:full_field_constraints:classifications}), the high-density peaks tracking the most massive structures also contain significant information capable of distinguishing between gravity models. A similar trend is observed for the F6 fiducial case. 

As noted in Section \ref{sec:method:likelihoods:truncPoisson}, we use an approximate truncated Poisson likelihood; however, adopting the approach of \cite{Kelly:2008wk} would allow for a more accurate likelihood and help mitigate the small bias seen in the F6 fiducial case. 

\begin{figure}
    \centering    
    \includegraphics[width=\linewidth]{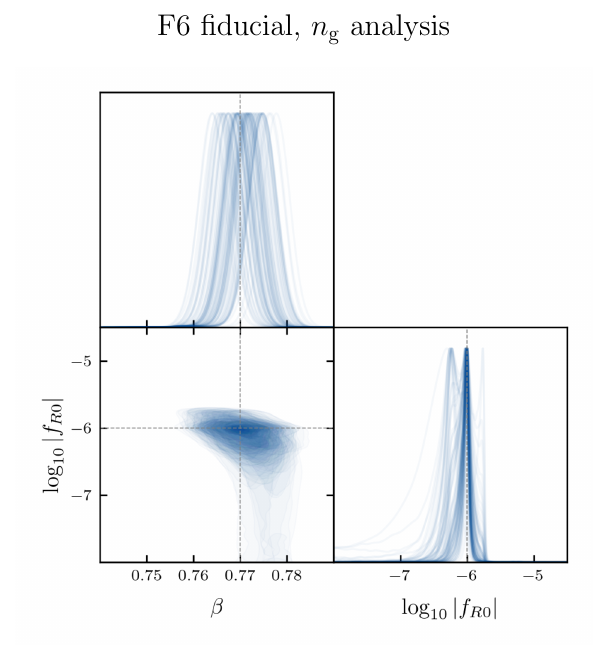}
    \caption{Constraint on $\beta$ and $|f_{R0}|$ from $n_{\rm{g}}$ full-field analysis for the F6 fiducial model across 100 Poisson realisations.  }
    \label{fig:varying_seeds_F6}
\end{figure}

\subsection{Different Poisson Seeds and DM Initial Conditions}\label{sec:robustness:Poisson_seeds}
We next assess the robustness of our pipeline against stochastic variations in Poisson noise. For this test, we generate a total of 100 independent noisy galaxy fields by varying the Poisson seed used to populate the $n_{\rm{g}}^{\rm{obs}}$ field, rerunning the full inference pipeline for each Poisson realisation. For this test, we impose a prior on $\beta \in [0.74, 0.79]$.

Figure~\ref{fig:varying_seeds_F6} shows the constraints obtained for the F6 fiducial model across these different realisations from our full-field analysis. While there are minor fluctuations in the exact position and size of the posterior contours due to the specific realisation of noise, the results remain remarkably consistent. The field-level analysis consistently outperforms the power spectrum analysis, recovering the true fiducial parameters with only slight variations across the different realisations.    

To quantify the statistical calibration of our method, we quantify how many of the recovered parameter constraints lie within $68\%$ of the fiducial value. We perform this test exclusively for the full-field $n_{\rm{g}}$ analysis in the F6 fiducial case, as we have clear expectations of the 68\% confidence levels for $|f_{R0}|$ and $\beta$, unlike in the GR fiducial case where we can only obtain an upper limit on $|f_{R0}|$. For these 100 realisations, we find that $66\%$ ($67\%$) of the recovered $\beta$ ($|f_{R0}|$) values lie within $1\sigma$ ($68\%$ confidence interval) of the fiducial value as expected. 
Overall, these results demonstrates that our findings are not driven by a fortunate choice of Poisson noise realisations. 

To assess the impact of cosmic variance on our parameter inference, we investigate the sensitivity of our inference framework to the underlying initial conditions of the COLA simulations. For this test, we generate five independent realisations of the dark matter density fields by varying the initial random seed used within COLA. For each realisation, we compute the full suite of 18 $\delta_{\rm{DM}}$ fields across our modified gravity grid (see Section~\ref{sec:method:param_estimation}) and re-run the full inference pipeline. We confirmed that our main conclusions, $i.e.$, the superior constraining power and degeneracy-breaking capability of the field-level approach, are not artifacts of a specific realisation of the cosmic density field. Ultimately, these different initial Gaussian random fields need to be constrained together with theoretical parameters when comparing with observed galaxy distributions \citep{Jasche_bayesian_2013}. 

\section{Conclusions and Discussions}\label{sec:conclusions}
In this work, we have developed and validated a field-level inference framework for simultaneously constraining galaxy bias and testing modified gravity models using the full 3D galaxy number counts field, $n_{\rm{g}}$. This approach exploits the non-Gaussian signatures that non-linear gravitational evolution imprints on smaller scales, features which are inaccessible to standard two-point statistics. By forward-modelling the non-linear matter field under various gravity strengths and galaxy bias, our field-level inference pipeline allows for robust joint parameter inference. We have validated this framework using fiducial mock datasets, representative of observed galaxies, for two scenarios representative of standard and modified gravity: GR and an $f(R)$ model with $\log_{10}|f_{R0}|=-6$. Our main conclusions can be summarised as follows:
\begin{itemize}
    \item Compared to a traditional power spectrum analysis, the full field-level likelihood provides significantly tighter constraints on both the strength of modified gravity, $|f_{R0}|$, and the galaxy bias parameter $\beta$. The F6 model is indistinguishable from GR at the 2$\sigma$ level when using only the power spectrum, while field-level inference allows the establishment of a lower bound that excludes GR at 2$\sigma$ confidence. The enhanced constraining power of the full number counts field arises from the inclusion of non-Gaussian information encoded in the Fourier amplitudes and phases.
    
    \item We have shown that, for fixed initial phases, we still get parameter constraints from phases alone even though those constraints are considerably weaker than the full field constraints. The phase  information is entirely lost in the power spectrum. 
    However, to obtain a tight constraint on both $\log_{10}|f_{R0}|$ and $\beta$, we must conduct an analysis using both the phases and amplitudes.

    \item We have shown through a cosmic web classification-based analysis that different cosmic environments (voids, walls, filaments, and clusters) provide complementary information into the origin of this additional constraining power. The field-level likelihood naturally combines information from all these environments.

    \item Specifically, we find that voxels with $n_{\rm{g}}^{\rm{obs}}=1$ are highly sensitive to increases in galaxy bias or modified gravity, while $n_{\rm{g}}^{\rm{obs}}\neq1$ drive the constraints distinguishing between lower gravity strength and galaxy bias. Further, we find that these $n_{\rm{g}}^{\rm{obs}}=1$ voxels sensitive to increased bias and gravity reside in the voids and walls of the underlying DM cosmic web.

    \item We demonstrate the robustness of our framework through extensive tests varying the initial conditions for structure formation and the random realisation of Poisson noise, and confirm that our results are not artefacts of any particular realisation. Additionally, we test the dependence of the field-level constraint on $n_{\rm{g}}$ thresholds.

\end{itemize}

Our results highlight the potential of field-level analyses to exploit the full information content of upcoming large-scale structure surveys like DESI, Euclid, and the Vera C. Rubin Observatory. Future work will include a comparative study between our field-level analysis and a combination of the power spectrum and bispectrum, including estimation of the bispectrum covariance. Our results also contribute to the broader, ongoing effort to understand the source of information gain in non-Gaussian cosmological fields when comparing field-level versus summary-statistic approaches \citep[e.g.,][]{Nguyen_et_al, 2024PhRvD.109d3526C, Akitsu_cosmology_2025, 2025JCAP...09..056S, Nikolac:2026ydg}. While the principled framework of \citet{2026arXiv260425385P} has provided a major stepping stone by rigorously demonstrating which components of field-level information map to specific summary statistics, the challenge of model misspecification remains an open concern \citep[e.g.,][]{2025JCAP...09..056S}. In this context, our analysis provides a complementary, spatially localized perspective. By isolating exactly which voxels within the cosmic web contribute most to our final parameter constraints, we offer a tangible, physical insight into precisely where the practical constraining power of FLI originates.

The important limitation of our analysis is that we fix the initial phases of the dark matter field. To apply our method to real data, it will be necessary to marginalise over these initial phases. This has been demonstrated within the BORG framework \citep{Jasche_bayesian_2013, Lavaux_unmasking_2016, Jasche_physical_2019, Stopyra:2023yqm}, which realized a digital twin of our Universe with the Manticore Project \citep{Manticore, Manticore_deep}. Information within the full field originates from both phase correlations and the non-Gaussian nature of the Fourier amplitudes. The result presented in Figure~\ref{fig:phase_only} indicates that the constraining power provided by phase information in isolation is notably weaker than the total field constraint. This implies that performing a phase marginalisation may not lead to a substantial information reduction for cosmological parameters, aligning with results from \cite{2023arXiv230404785P}. It is yet to be determined, however, if such a result is applicable to modified gravity constraints extracted from galaxy number density fields.

Although COLA simulations are significantly faster than full N-body simulations, they remain too computationally expensive for direct use in inference in modified gravity models that require accurate treatment of screening mechanisms. Field-level emulators have been developed to address this issue \citep{Jamieson:2024fsp, Saadeh:2024vuj, Saadeh:2025gnz}, and our pipeline can readily incorporate such emulators. 

Our analysis setup is based on \cite{Jasche_physical_2019} and adopts a local matter-density bias model. At the voxel sizes considered in this work, this bias prescription is unlikely to be sufficiently accurate, and more sophisticated bias models are required. Improved modelling of the stochastic amplitude of the galaxy field is required for both power-spectrum analyses and FLI \citep{Britt:2024knf}. 
Furthermore, applying the pipeline to real survey data will require a careful treatment of observational systematics. Since the original BORG framework on which our analysis is based, substantial progress has been made in field-level inference. For example, recent developments within the Manticore project~\citep{Manticore,Manticore_deep} have introduced improved treatments designed to mitigate model misspecification. We plan to update our analysis in future work by incorporating these advances, with the ultimate goal of enabling robust applications of our method to real observational data.

Finally, we intend to extend this framework to include additional cosmological parameters ($e.g.$, $\sigma_8$) and to explore a broader range of modified gravity models. Furthermore, extending this framework to incorporate weak lensing provides a complementary probe of the underlying DM distribution. Integrating weak lensing predictions with our current galaxy number counts field offers the opportunity to break degeneracies between galaxy bias and cosmological parameters more robustly. Weak lensing map generation in modified gravity has already been implemented in COLA \citep{Hoyland_fast_2025}, while the feasibility of field-level inference of cosmic shear has been successfully demonstrated in \cite{2023arXiv230404785P}. Simultaneously, incorporating redshift-space distortions into the forward-modeling pipeline offers another fruitful avenue for lifting these bias degeneracies at the map level \citep{2023JCAP...10..069S}. We explore the possibility of combining these complementary field-level approaches in future work.

\section*{Acknowledgements}
SH is supported by the UK Science and Technology Facilities Council (STFC) grant number ST/X508688/1 and funding from the University of Portsmouth. DS and KK were supported by STFC grant number ST/W001225/1. KK is supported also by STFC grant number ST/B001175/1. DS is additionally supported by the European Research Council (ERC) Advanced Grant UNCA (UKRI Frontiers Research Guarantee No.~EP/Z533877/1). HD is supported by a Royal Society University Research Fellowship (grant no. 211046).

Numerical computations were carried out on the \textsc{Sciama} High Performance Computing (HPC) cluster, which is supported by the Institute of Cosmology and Gravitation, the South-East Physics Network (SEPNet) and the University of Portsmouth.

For the purpose of open access, we have applied a Creative Commons Attribution (CC BY) licence to any Author Accepted Manuscript version arising.

\section*{Data Availability}

The supporting research data and code are available upon reasonable request from the corresponding author.


\bibliographystyle{mnras}
\bibliography{references}

\appendix




\bsp	
\label{lastpage}

\end{document}